\documentclass{svjour3arxiv}
\smartqed  

\usepackage{graphicx}
\usepackage{lineno,hyperref,bm,float, soul, amsfonts, amsmath, siunitx}
\usepackage{xcolor}
\modulolinenumbers[5]
\newcommand{\T}{^{\mbox{\tiny T}}}
\newcommand{\Ts}{^{\,\mbox{\tiny T}}}
\newcommand{\B}[1]{{\bm #1}}
\newcommand{\U}[1]{\hat{\bm #1}}
\newcommand{\dd}{\; \text{d}}

\newcommand{\X}{\mathbb{X}}

\begin{document}

\title{Fast 2-impulse non-Keplerian orbit-transfer using the Theory of Functional Connections\footnote{A previous version was presented as: De Almeida Junior, A.K., Johnston, H., Leake, C., and Mortari, D. ``Evaluation of transfer costs in the Earth/Moon system using the Theory of Functional Connections,'' AAS 20-596, Astrodynamics Specialist Conference, August 9-12, Lake Tahoe, CA.}}

\author{Allan K. de Almeida Junior \and Hunter Johnston \and Carl Leake \and Daniele Mortari}

\institute{Allan K. de Almeida Junior \at
	INPE - National Institute for Space Research, S\~ao Jos\'e dos Campos, SP, Brazil\\
	Aerospace Engineering, Texas A\&M University, College Station, TX.\\ \email{allan.junior@inpe.br}
	\and
	Hunter Johnston \at
	Aerospace Engineering, Texas A\&M University, College Station, TX.
	\email{hunterjohnston@tamu.edu}
	\and
	Carl Leake \at
	Aerospace Engineering, Texas A\&M University, College Station, TX.
	\email{leakec@tamu.edu}
	\and
	Daniele Mortari \at
	Aerospace Engineering, Texas A\&M University, College Station, TX.
	\email{mortari@tamu.edu}
}

\date{}

\maketitle

\begin{abstract}
This study applies a new approach, the Theory of Functional Connections (TFC), to solve the two-point boundary-value problem (TPBVP) in non-Keplerian orbit transfer. The perturbations considered are drag, solar radiation pressure, higher-order gravitational potential harmonic terms, and multiple bodies. The proposed approach is applied to Earth-to-Moon transfers, and obtains exact boundary condition satisfaction and with very fast convergence. Thanks to this highly efficient approach, perturbed pork-chop plots of Earth-to-Moon transfers are generated, and individual analyses on the transfers' parameters are easily done at low computational costs. The minimum fuel analysis is provided in terms of the time of flight, thrust application points, and relative geometry of the Moon and Sun. The transfer costs obtained are in agreement with the literature's best solutions, and in some cases are even slightly better.
\keywords{
      {Two point boundary value problem}{}   \and
      {Transfers in space}{} \and
      {Bi-circular four body problem}{} \and {Theory of Functional Connections}{}
     }
\end{abstract}
	
\section{Introduction}
	
The main problem in evaluating perturbed orbit transfer costs is solving a two-point boundary-value problem (TPBVP) on a nonlinear dynamics. This problem is traditionally solved using shooting methods. This paper focuses its attention on Earth-to-Moon co-planar orbit transfer as an example of TPBVP orbit transfer by also including Sun gravitational perturbation.
	
The three-body problem (3BP) \cite{poincare1890equations,DIACU1999175,HOLMES1990137} has traditionally been used as an initial guess to solve multi-body astrodynamics problems. 
Earth/Moon transfers have recently attracted the interest of the scientific community following the work done by \cite{sweetser1991estimate}. 
This transfer can be done through the use of low thrust maneuvers \cite{2018palau,doi:10.2514/2.4210,doi:10.2514/3.21515} or two-impulse thrust, a Hohmann maneuver, solved via the patched restricted three-body problem \cite{da2012optimal,5586384}, full ephemeris $n$-body problem \cite{doi:10.2514/3.21079,LEI2013917}, restricted three body-problem \cite{pernicka95,YAGASAKI2004313,topber05,doi:10.2514/1.7702,da2012optimal}, restricted four-body problem \cite{ASSADIAN2010398}, and bi-circular restricted four-body problem \cite{1992sfm..proc.1113Y,Yagasaki2004,mingtop2011,dasilvamar2011,doi:10.2514/1.55426,mingotti2012efficient,topputo2013optimal,Oshima2017,Onozaki2017,QI2017106,oshimatop19}. 
	
The Earth/Moon transfer TPBVP has also been solved using a gradient method based on the Lambert problem as an initial guess \cite{pradobroucke}, a direct transcription and multiple shooting method \cite{topputo2013optimal}, or a more recently developed method based on the gradient method and Jacobi integral variational equation \cite{GAGGFILHO2019312}.
	
In general, most of the techniques used to solve this problem in the mentioned references, such as the shooting method, are computationally intensive, because they need to integrate the differential equations (dynamics) multiple times, because the shooting method iterates the initial conditions to match the boundary conditions. In this study, the authors adopt the recently developed \textit{Theory of Functional Connections} (TFC) \cite{U-TFC,LDE,NDE,M-TFC,M-TFC-new} to solve the TPBVP. This method provides functionals that analytically satisfy the boundary conditions. This way, the initial constraint optimization problem becomes unconstrained, and a single unconstrained nonlinear least-squares approach can be applied to solve the optimization problem. In contrast, previous methods need to solve the problem multiple times. 
	
The efficiency also involves the accuracy in terms of position and velocity errors. In particular, the code uses automatic differentiation and a just-in-time (JIT) compiler \cite{JaxGithub,JaxOriginalPaper} to implement the nonlinear least-squares. As a result, the run time to find a solution for a given set of parameters is on the order of seconds.
This fast method for solving the TPBVP with nonlinear differential equations has facilitated the generation of multi-body, perturbed, orbit transfer pork-chop plots and individual analyses on the transfer parameters.
	
The two-impulse transfer is done using the circular co-planar restricted 3BP and bi-circular, bi-planar four-body problem by including the Sun. A set of four parameters of this transfer are investigated to identify the optimal combination that minimizes fuel consumption. These parameters are 1) first impulse application point on the departure LEO, 2) second impulse application point on the arrival orbit around the Moon, 3) time of flight, and 4) Sun relative direction with respect to the Earth/Moon system. In particular, the transfer time considered is short, up to seven days.
	
This article is organized as follows. The mathematical background, the definition of the problem, and the models adopted are shown (next section). In the results section, the parameters' influence over the equivalent fuel cost are analyzed for a transfer between a LEO and another circular orbit close to the Moon. Such results are compared with those available in the literature, and the major findings are summarized in the conclusions section.
	
\section{Mathematical background}
	
In this section, the problem is defined, and the mathematical models and tools are explained. First, a recent technique to solve the two-point boundary value problems for a system of differential equations is summarized. Then, the problem is defined, and the mathematical model used to evaluate the transfer costs is presented.
	
\subsection{Summary of the Theory of Functional Connections}
	
The numerical technique to solve differential equations based on the Theory of Functional Connections (TFC) \cite{U-TFC,M-TFC} has been shown to produce solutions with machine-level error within milliseconds for both linear \cite{LDE} and nonlinear \cite{NDE} differential equations and for a wide array of constraint cases. Several works have utilized the TFC framework including, hybrid systems \cite{hybrid_tfc}, optimal control problems \cite{EOL_EOI,FOL}, quadratic and nonlinear programming \cite{QP_NLP}, and other applications \cite{Selected,DeepTfc,SVM}.
	
This approach's foundation is the ability to derive a class of functionals, called constrained expressions, which have the characteristic of always analytically satisfying assigned linear constraints. Using these constrained expressions, the differential equation can be transformed into an algebraic expression that can ultimately be solved via a variety of optimization techniques. In general, the TFC methodology provides \emph{functional interpolation} by embedding a set of $n$ linear constraints. The general expression from \cite{M-TFC-new} can be adopted for vector equations using the following functional,
\begin{equation}\label{eq:generalCE}
		x (t, g (t)) = g (t) + \sum_{j = 1}^k \phi_j (t) \, \rho_j(t,g(t)), 
\end{equation}
where $g(t)$ is a free function and $\rho_j(t,g(t))$ are called projection functionals since they project the free function onto the space of functions that satisfy the constraints. They are derived by setting the constraint function equal to zero and replacing $x(t)$ with $g(t)$. In addition, $\phi_j(t)$ are called switching functions; by definition, they are equal to $1$ when evaluated at the constraint they are referencing and equal to $0$ when evaluated at all other constraints. The switching functions are composed of a set of linearly independent functions called support functions, $s_k(t)$, with unknown coefficients $\alpha_{ij}$, such that $\phi_j (t) = s_i (t) \, \alpha_{ij}$. The following step-by-step procedure can be used to derive the switching functions:
\begin{enumerate}
		\item Choose $k$ support functions, $s_k(x)$;
		\item Write each switching function as a linear combination of the support functions with unknown coefficients;
		\item Based on the switching function definition, write a system of equations to solve the unknown coefficients.
\end{enumerate}
Once the switching functions and projection functionals have been constructed, they can be substituted into Eq. \eqref{eq:generalCE} to form the constrained expression. For the interested reader, this process is described in far more detail with examples in \cite{M-TFC-new}.
	
\subsection{Application of the Theory of Functional Connections to solve two-point boundary value problems}
	
For convenience, a step-by-step derivation of the constrained expression used in the paper is presented below. Additionally, a summary of the numerical technique is also provided. Since we deal with a bi-planar, bi-circular four-body problem, a two-dimensional vector is defined as $\X = \{x, y\}\T$, where $\X_1 = x$ and $\X_2 = y$. This problem's dynamics are governed by a nonlinear second-order vectorial differential equation subject to boundary value and derivative constraints. In general, a differential equation of this kind can be expressed as
	\begin{equation}\label{eq:genODE}
		F_i(t, \X_j, \dot{\X}_j, \ddot{\X}_j) = 0 \quad \text{subject to:} \quad \begin{cases} X_j(t_0) = X_{0_j}\\ 
			X_j(t_f) = X_{f_j}\end{cases} \quad \text{where} \quad i,j = 1,2.
	\end{equation}
Since the constraints for each component of the vector $\X_i$ are of the same form, Eq. \eqref{eq:generalCE} can easily be adapted to the vectorial form
\begin{equation}\label{eq:generalCE_vector}
		\X_i (t, g_i (t)) = g_i (t) + \sum_{j = 1}^k \phi_j (t) \, \rho_j(t,g_i(t)).
\end{equation}
Following the process detailed in the previous section, one must first choose the support functions $s_k(t)$. For these specific constraints, the support functions are defined in the same manner as \cite{FOL}, where $s_k (t) = t^{(k-1)}$. Now, the definition of the switching functions is used to produce a set of equations. For example, the first switching function has the two equations,
\begin{equation*}
		\phi_1(t_0) = 1 \quad \text{and} \quad \phi_1(t_f) = 0.
\end{equation*}
These equations are expanded in terms of the support functions,
\begin{align*}
		\phi_1(t_0) &= (1) \cdot \alpha_{11} + (t_0) \cdot\alpha_{21} = 1 \\
		\phi_1(t_f) &= (1) \cdot \alpha_{11} + (t_f) \cdot\alpha_{21} = 0
	\end{align*}
	which can be compactly written as
	\begin{equation*}
		\begin{bmatrix} 1 & t_0 \\ 1 & t_f  \end{bmatrix} \begin{Bmatrix} \alpha_{11} \\ \alpha_{21} \end{Bmatrix} = \begin{Bmatrix} 1 \\ 0  \end{Bmatrix}.
	\end{equation*}
	The same is done for the other switching function to produce a set of equations that can be solved by matrix inversion,
	\begin{align*}
		\begin{bmatrix} 1 & t_0  \\ 1 & t_f  \end{bmatrix} \begin{bmatrix} \alpha_{11} & \alpha_{12} \\ \alpha_{21} & \alpha_{22} \end{bmatrix} &= \begin{bmatrix} 1 & 0 \\ 0 & 1 \end{bmatrix} \\
		\begin{bmatrix} \alpha_{11} & \alpha_{12} \\ \alpha_{21} & \alpha_{22} \end{bmatrix} &= \begin{bmatrix} 1 & t_0 \\ 1 & t_f \end{bmatrix}^{-1} = 
		\begingroup
		\renewcommand*{\arraystretch}{2}
		\begin{bmatrix} \dfrac{t_f}{t_f - t_0} & \dfrac{-t_0}{t_f - t_0} \\ \dfrac{-1}{t_f - t_0} & \dfrac{1}{t_f - t_0} \end{bmatrix},
		\endgroup
	\end{align*}
	leading to the switching functions,
	\begin{equation*}
		\phi_1(t) = \frac{t_f - t}{t_f - t_0} \quad \text{and} \quad
		\phi_2(t) = \frac{t - t_0}{t_f - t_0}.
	\end{equation*}
	From their definition, the projection functionals are
	\begin{equation*}
		\rho_1(t,g_i(t)) = \X_{i_0} - g_i(t_0) \quad \text{and} \quad
		\rho_2(t,g_i(t)) = \X_{i_f} - g_i(t_f).
	\end{equation*}
	Collecting these terms in the form of Eq. \eqref{eq:generalCE_vector}, the constrained expression can be expressed as
	\begin{align*}
		\X_i (t, g_i (t)) &= g_i (t) + \frac{t_f - t}{t_f - t_0} \Big(\X_{i_0} - g_i(t_0)\Big) + \frac{t - t_0}{t_f - t_0} \Big(\X_{i_f} - g_i(t_f)\Big).
	\end{align*} 
Now, substituting the constrained expression into the differential equation, Eq. \eqref{eq:genODE}, the differential equation is transformed into one subject to no constraints,
\begin{equation}\label{eq:tildeF}
		\tilde{F}_i(t, g_j(t), \dot{g}_j(t), \ddot{g_j}(t)) = 0 \quad i,j = 1,2.
\end{equation}
In order to solve this new equation, three major steps must be taken: 1) define the free-function $g_i(t)$, 2) discretize the domain, and 3) solve the resulting algebraic equation. The following text summarizes this process.
	
\subsubsection{Definition of the free-function \texorpdfstring{$g_j(t)$}{g(t)}:} 

For this application, let the free-function be expressed as a linear combination of basis functions multiplied with unknown coefficients according to
\begin{equation}\label{eq:general_multi_g}
		g_i(t) =  \B{h}\T \B{\xi}_i,
\end{equation}
where $\B{h}\in\mathbb{R}^m$ is a vector of the basis functions and $\B{\xi}_i\in\mathbb{R}^m$ is the unknown coefficient vector associated with the $i^{\text{th}}$ vector component. In general, the basis functions that make up $\B{h}$ will not coincide with the problem domain, e.g., Chebyshev and Legendre polynomials are defined in $[-1,1]$. Therefore, let the basis functions be defined for $z \in [z_0, z_f]$ and the problem be defined for $t \in [t_0, t_f]$. In order to use the basis functions, a map between the basis function domain, $z$, and problem domain, $t$, must be created. The simplest map is linear,
\begin{equation}\label{eq:linearMapping}
		z = z_0 + \frac{z_f-z_0}{t_f-t_0}(t - t_0) \quad \longleftrightarrow \quad t = t_0 + \frac{t_f-t_0}{z_f-z_0}(z - z_0).
\end{equation}
Using this map, the subsequent derivatives of the free function can be computed as
\begin{equation*}
		\frac{\dd^n g_i}{\dd t^n} = \left(\frac{\dd z}{\dd t}\right)^{n}  \frac{\dd^n \B{h}\Ts}{\dd z^n} \B{\xi}_i,
\end{equation*}
where, by defining
\begin{equation*}
	c := \frac{\dd z}{\dd t} = \frac{z_f - z_0}{t_f - t_0},
\end{equation*}
the derivative computations can be simply written as
\begin{equation*}
	\frac{\dd^n g_i}{\dd t^n} = c^{n} \frac{\dd^n \B{h}\Ts}{\dd z^n} \B{\xi}_i,
\end{equation*}
where $c$ is referred to as the mapping coefficient.
	
\subsubsection{Domain discretization:}

In order to solve problems numerically, the problem domain, and therefore the basis function domain, must be discretized. Since this article uses Chebyshev orthogonal polynomials, an optimal discretization scheme is the Chebyshev-Gauss-Lobatto nodes \cite{Collo1,Collo2}. For, $N+1$ points, the discrete points are calculated as,
\begin{equation}\label{eq:collo}
	z_k = -\cos\left(\frac{k \pi}{N}\right) \quad \text{for} \quad k = 0, 1, 2, \cdots, N.
\end{equation}
When compared with the uniform distribution, the above collocation point distribution results in a much easier condition of the matrix to be inverted in the least-squares as the number of basis functions, $m$, increases. The collocation points can be realized in the problem domain through the relationship provided in Eq. \eqref{eq:linearMapping}.
	
\subsubsection{Solution of the resulting algebraic equation:}\label{sec:steps}
By defining $g_i(t)$ according to Eq. \eqref{eq:general_multi_g}, our definition of Eq. \eqref{eq:tildeF} is converted into
\begin{equation}\label{eq:finalSystem}
	\tilde{F}_i(t, \B{\xi}_j) = 0 \quad i,j = 1,2.
\end{equation}
	
Now, if the original differential equation \eqref{eq:genODE} is linear, then Eq. \eqref{eq:finalSystem} is linear in the unknown $\B{\xi}_j$ coefficients and can be solved with any least-squares technique. However, if the original differential equation is nonlinear, as is the governing differential equation of this paper, a nonlinear least-squares approach must be used. First, using Eq. \eqref{eq:finalSystem} one must construct the loss function of the system, which takes the form,
\begin{equation*}
	\mathbb{L}(\Xi) = \begin{Bmatrix} \tilde{F}_1(t_0, \Xi) \\ \vdots \\ \tilde{F}_1(t_f, \Xi) \\ \tilde{F}_2(t_0, \Xi) \\ \vdots \\ \tilde{F}_2(t_f, \Xi) \end{Bmatrix} = \B{0}
\end{equation*} 
where $\Xi = \{\B{\xi}_1\T \text{  } \B{\xi}_2\T\}\T$ and the function $\tilde{F}_i(t,\B{\xi}_j)$ is evaluated at the Chebyshev-Gauss-Lobatto nodes through Eqs. \eqref{eq:linearMapping} and \eqref{eq:collo}. In order to solve for the unknown coefficients, the parameter update equation is provided by,
\begin{equation}\label{eq:lsstep}
	\Xi_{k+1} = \Xi_k + \Delta \Xi_k
\end{equation}
where the $\Delta \Xi_k$ can be defined using classic least-squares,
\begin{equation*}
	\Delta \Xi_k = -\Big(\mathbb{J}\T(\Xi_k) \mathbb{J}(\Xi_k) \Big)^{-1} \mathbb{J}\T(\Xi_k) \mathbb{L}(\Xi_k)
\end{equation*}
or in this paper through a QR decomposition method. It follows that the Jacobian matrix is,
\begin{equation*}
	\mathbb{J}(\Xi) = \begin{bmatrix} \frac{\partial \tilde{F}_1(t_0, \Xi)}{\partial \B{\xi}_1} & \frac{\partial \tilde{F}_1(t_0, \Xi)}{\partial \B{\xi}_2} \\ \vdots & \vdots \\ \frac{\partial \tilde{F}_1(t_f, \Xi)}{\partial \B{\xi}_1} & \frac{\partial \tilde{F}_1(t_f, \Xi)}{\partial \B{\xi}_2} \\  \frac{\partial \tilde{F}_2(t_0, \Xi)}{\partial \B{\xi}_1} & \frac{\partial \tilde{F}_2(t_0, \Xi)}{\partial \B{\xi}_2} \\ \vdots & \vdots \\ \frac{\partial \tilde{F}_2(t_f, \Xi)}{\partial \B{\xi}_1} & \frac{\partial \tilde{F}_2(t_f, \Xi)}{\partial \B{\xi}_2} \end{bmatrix}.
\end{equation*}
	
\subsection{The bi-planar,  bi-circular four-body problem}
	
A vector $\B{q}$ is defined such that it locates the origin of a rotating frame of reference with respect to an inertial one. The equation of motion of a particle in this rotating frame of reference is \cite{symon} (chapter 7)
\begin{equation}\label{eq:rot1}
	\frac{\dd^2 \B{r}}{\dd t^2} + 2 \, \B{\omega} \times \frac{\dd \B{r}}{\dd t} + \B{\omega} \times \left(\B{\omega} \times \B{r}\right) +\frac{\dd^{*} \B{\omega}}{\dd t} \times \B{r} + \frac{\dd^{*2} \B{q}}{\dd t^2} = \B{a},
\end{equation}
where $\B{r}$ locates the spacecraft in the rotating frame, $\B{\omega}$ is the angular velocity of the rotating frame, and $\B{a}$ is the specific force acting on the particle. In this equation, the derivatives of the fourth and fifth members of the left side must be taken with respect to the time in the inertial frame, while the others are taken in the rotating frame - not taking into consideration the motion of their bases with respect to the inertial frame.
	
\begin{figure*}
	\centering\includegraphics[width=0.6\linewidth]{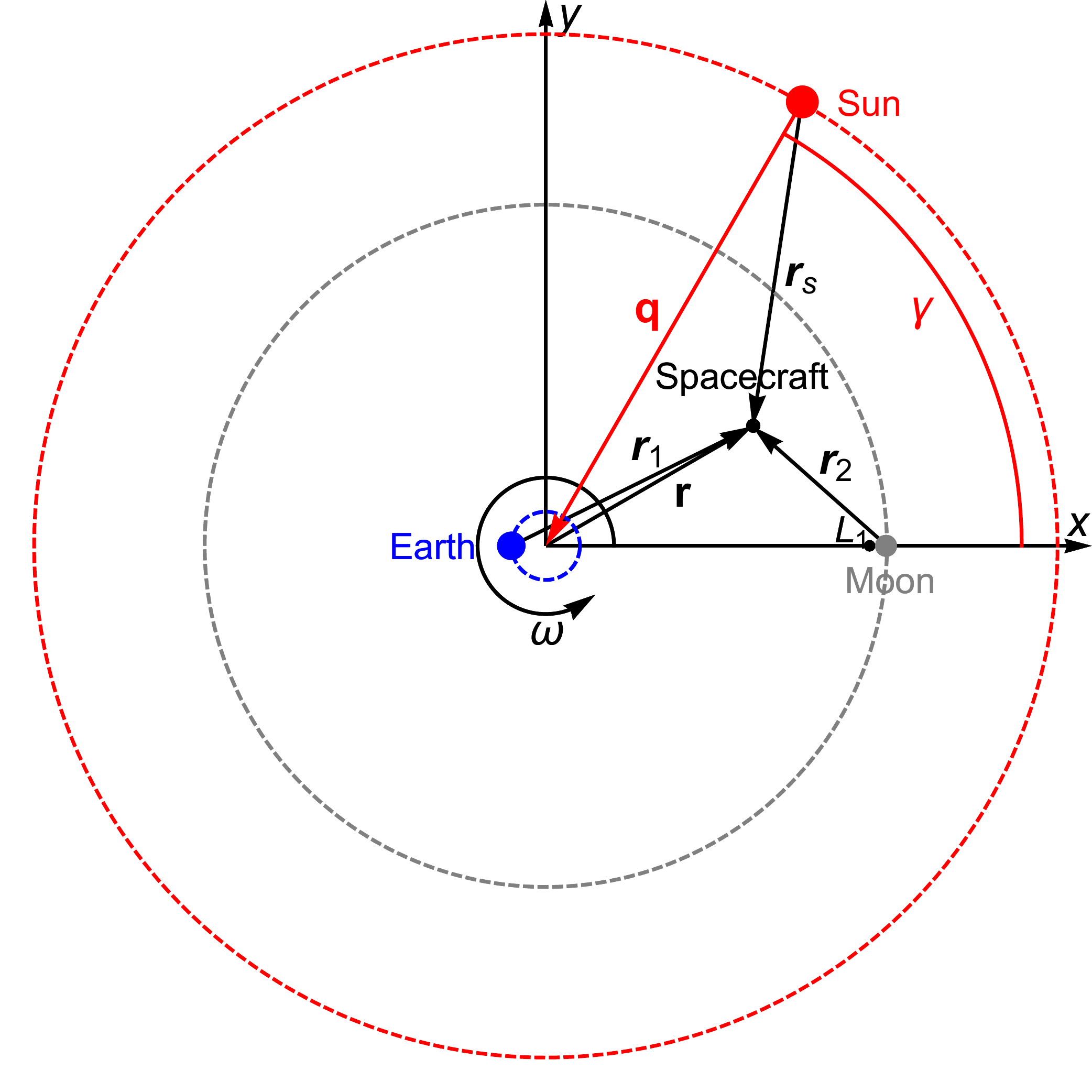}
	\caption{The Earth/Moon rotating frame of reference.}
	\label{fig:f1}
\end{figure*}

In the bi-planar, bi-circular four-body problem, the Earth and the Moon rotate in a circular orbit around their barycenter, which in turn rotates in a circular orbit around the Sun. Moreover, the two orbital planes coincide with each other. The rotating frame of reference can be seen in Fig. \ref{fig:f1}.
In our problem, the center of the inertial frame of reference coincides with the Sun's position. The vector $\B{q}$ locates the barycenter of the Earth and the Moon, which are located along the $x$ axis of the rotating frame of reference. The position of the origin of the rotating frame (the barycenter of the Earth/Moon system) written in the inertial frame is $\B{q}=\{R_s \cos{\theta_1},R_s \sin{\theta_1}\}\T$, where $R_s$ is the distance between the Sun and the barycenter of the pair Earth/Moon and $\theta_1=(\omega_r t+\gamma-\pi)$ is the angle that locates the center of the rotating frame with respect to the positive side of the $x$-axis of the inertial frame, where $(\gamma-\pi)$ is an initial phase and $\omega_r$ is the constant angular velocity of the barycenter of the Earth/Moon system in the inertial frame of reference. The acceleration of this barycenter with respect to the inertial frame written in the rotating frame of reference is
\begin{equation}\label{eq:0}
	\dfrac{\dd^{*2} \B{q}}{\dd t^2} = \dfrac{\mu_s}{R_s^2} \begin{Bmatrix} \cos{(\omega_s t + \gamma)} \\ \sin{(\omega_s t + \gamma)}\end{Bmatrix},
\end{equation}
where $\mu_s$ is the gravitational parameter of the Sun and $\omega_s=\omega_r-\omega$. The distance of the Earth and the Moon from their center of mass is given by $d_1 = R  \mu_2/(\mu_1 + \mu_2)$ and $d_2 = R  \mu_1 /(\mu_1 + \mu_2)$, respectively, where $\mu_1$ and $\mu_2$ are the gravitational parameters of the Earth and the Moon, respectively, and $R = d_1 + d_2$ is the total distance between them. The position vectors of the spacecraft relative to the Earth ($\B{r}_1$) and the Moon ($\B{r}_2$) can be written as $\B{r}_1 = \B{r} + d_1 \, \U{i}$ and $\B{r}_2 = \B{r} - d_2 \, \U{i}$,	where $[\U{i}, \U{j}]$ are unit vectors along the $[x, y]$ axes, respectively. Their magnitudes are $r_1 = \|\B{r}_1\|$ and $r_2 = \|\B{r}_2\|$. Similarly, the position vector of the spacecraft relative to the Sun is given by $\B{r}_s=\B{r}-\B{p}_s$, where $\B{p}_s=R_s\{ \cos{(\omega_s t+\gamma)},\sin{(\omega_s t+\gamma)}\}\T$ locates the Sun in the rotating frame of reference.
	
Hence, the equation of motion, Eq. (\ref{eq:rot1}), of a spacecraft under the gravitational influence of the Earth, the Moon, and the Sun written in the rotating frame of reference is
\begin{eqnarray}\label{eq:1}
	\frac{\dd^2 \B{r}}{\dd t^2} + 2 \, \B{\omega} \times \frac{\dd \B{r}}{\dd t} + \B{\omega} \times \left(\B{\omega} \times \B{r}\right) = - \dfrac{\mu_1}{r_1^3} \, \B{r}_1 - \dfrac{\mu_2}{r_2^3} \, \B{r}_2\\ \nonumber
	-\frac{\mu_s}{r_s^3}\B{r}_s-\frac{\mu_s}{R_s^2}\{\cos{(\omega_s t+\gamma)}, \sin{(\omega_s t+\gamma)}\}\T.
\end{eqnarray}
	
In the case where $\mu_s=0$, Eq. \eqref{eq:1} is reduced to the planar circular restricted three-body problem. This simpler case will be investigated first in section \ref{sec:results}. Afterwards, the bi-circular, bi-planar restricted four-body problem will be used to model the spacecraft's motion.

\subsection{The equivalent fuel cost \texorpdfstring{$\Delta V$}{DV}}
	
The initial and final positions of the transfer are denoted with $A$ and $B$, respectively. A first impulse is applied to the spacecraft at the point $A$ such that it travels to the point $B$ during a time of flight $T$ under the gravitational influence of the main bodies. The initial and final velocities ($\B{V}_A$ and $\B{V}_B$, respectively) of the boundary value problem in the rotating frame are evaluated through the use of the TFC.	The equivalent fuel cost $\Delta V$ is the sum of the magnitude of the initial and the final impulses, and hence it is given by
\begin{equation*}
	\Delta V=\|\delta \B{V}_A\|+\|\delta \B{V}_B\|,
\end{equation*}
where $\|\delta \B{V}_A\|$, the magnitude of the difference between the initial velocity and the one that the spacecraft had just before the impulse, is given by
\begin{equation*}
	\|\delta \B{V}_A\|=\|\B{V}_A-\B{V}_{Ai}\|,
\end{equation*}
and $\|\delta \B{V}_B\|$, the magnitude of the difference between the final velocity and the velocity required for the spacecraft to achieve its final orbit, is similarly given by
\begin{equation*}
	\|\delta \B{V}_B\|=\|\B{V}_{Bf}-\B{V}_B\|.
\end{equation*}
	
\subsection{Transfer from an initial circular orbit around the Earth to a final circular orbit around the Moon}\label{sec:transfer}
	
\begin{figure}[!ht]
	\centering
	\includegraphics[width=\linewidth]{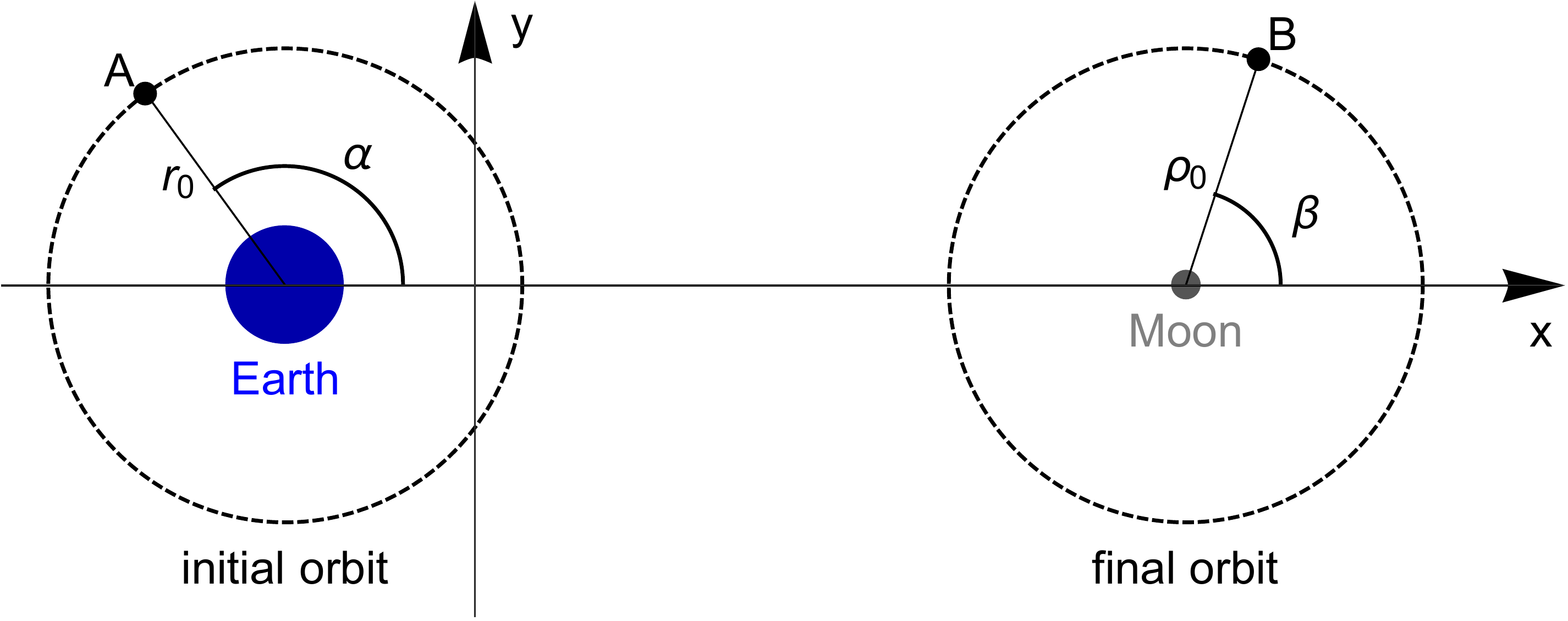}
	\caption{The initial (A) and final (B) points of the transfer.}
	\label{fig:g2}
\end{figure}

A fixed inertial frame of reference is defined such that its center coincides with the center of the Earth. The spacecraft is in a circular orbit of radius $r_0$ around the Earth with an angular velocity $\omega_0=\sqrt{\mu_1/r_0^3}$ in this inertial frame of reference. A second inertial frame of reference is also defined such that its center coincides with the center of the Moon. The spacecraft will be inserted in a second orbit around the Moon of radius $\rho_0$ and angular velocity $\Omega_0 = \pm\sqrt{\mu_2/\rho_0^3}$ in this inertial frame. The first impulse is applied when the spacecraft's initial position vector relative to the Earth in the circular orbit makes an angle $\alpha$ with the $x$-axis. The second impulse is applied when the spacecraft crosses the final circular orbit around the Moon at the point $B$, at which point the orbit makes an angle $\beta$ with the $x$-axis in the rotating frame of reference.
Both points, $A$ and $B$, are shown in Fig. \ref{fig:g2}.
	
The position vector at the initial time of motion in the rotating frame of reference is
\begin{equation*}
	\B{r}_{A}=\begin{Bmatrix}-d_1+r_0 \cos \alpha\\r_0 \sin \alpha\end{Bmatrix}.
\end{equation*}
The initial velocity before the impulse in the rotating frame of reference is 
\begin{equation*}
	\B{V}_{Ai} =  \begin{Bmatrix} -(\omega_0-\omega) r_0 \sin \alpha \\ (\omega_0-\omega) r_0 \cos \alpha\end{Bmatrix}.
\end{equation*}
The position vector at the final time of motion in the rotating frame is
\begin{equation*}
	\B{r}_{B}= \begin{Bmatrix} d_2+\rho_0 \cos \beta \\ \rho_0 \sin \beta\end{Bmatrix}
\end{equation*}
and the final velocity of the circular orbit after the second impulse (at $B$) in the rotating frame is
\begin{equation*}
	\B{V}_{Bf}=\begin{Bmatrix}-(\Omega_0-\omega) \rho_0 \sin \beta\\(\Omega_0-\omega) \rho_0 \cos \beta\end{Bmatrix}.
\end{equation*}
In the case of a counter-clockwise final orbit around the Moon, the angular velocity is $\Omega_0=\sqrt{\mu_2/\rho_0^3}$. In the case the orbit is clockwise, the angular velocity is $\Omega_0 = -\sqrt{\mu_2/\rho_0^3}$.

\subsection{Error estimation}
	
The initial and final positions are known, and the initial and final velocities are obtained through the TFC method. The initial position and initial velocity form a set of initial conditions.
This set of initial conditions is numerically integrated for the same system via a fourth-order Runge-Kutta integrator. A position error $P_E$ is defined as the magnitude of the difference between the final position (the point $B$) and the position given by the Runge-Kutta integrator ran for the corresponding time of travel $T$. Analogously, a velocity error $V_E$ is defined as the magnitude of the difference between the velocity at the final time of motion $t_f$ obtained by the solution via TFC and the Runge-Kutta method using the initial conditions given by the TFC solution. Note that $P_E$ and $V_E$ are not true errors since they also accumulate numerical errors generated by the RK integrator. On the other hand, there is no true solution available for this system, and such a combination of errors may give us a good estimation of the upper limits of the true errors for this model.
	
\subsection{Values of the parameters}
	
The values of the parameters for the Earth/Moon system used in this research are given in Table \ref{tab:1}. To facilitate comparisons, these values are the same as those used by \cite{simo1995book,YAGASAKI2004313,Yagasaki2004,topputo2013optimal}.
\begin{table}[ht]
	\caption{Values of the parameters for the Earth/Moon system \cite{simo1995book}.}
	\centering
	\begin{tabular}{cc}
		\hline
		$R$&$3.84405000\times10^{8}~\text{m}$\\[0.5ex]
		$R_s$&$1.49460947424915\times 10^{11}~\text{m}$ \\[0.5ex]
		$\mu_1$&$3.975837768911438\times10^{14}~\text{m}^3/\text{s}^2$ \\[0.5ex]
		$\mu_2$&$4.890329364450684\times10^{12}~\text{m}^3/\text{s}^2$ \\[0.5ex]
		$\mu_s $&$ 1.3237395128595653 \times 10^{20}~\text{m}^3/\text{s}^2$ \\[0.5ex] 
		$\omega$&$2.66186135\times 10^{-6}~\text{1/s}$ \\[0.5ex]
		$\omega_s$&$-2.462743433827215\times 10^{-6}~\text{1/s}$ \\[0.5ex]
		Radius of the Earth&$6.378\times 10^{6}~\text{m}$ \\[0.5ex]
		Radius of the Moon&$1.738\times 10^{6}~\text{m}$ \\[0.5ex]
		\hline
	\end{tabular}
	\label{tab:1}
\end{table}
	
\section{Results}\label{sec:results}
		
The results for the transfer are presented in this section. The spacecraft departs from an initial circular orbit around the Earth with an altitude of $167$ km and will be inserted into a final orbit around the Moon with an altitude of $100$ km, according to Fig. \ref{fig:g2}.
	
A spiral is used as an initial guess to find the possible solutions of the system. The spiral is defined as a linear variation on time of the radius and the angle in polar coordinates, from the initial position to the final one.	
	
Each iteration of Eq.~(\ref{eq:lsstep}) takes less than a tenth of a second to be evaluated (usually much less) for a residual on the order of $10^{-15}$. For most cases, a solution is found between 10 to 20 iterations. The cases where no convergence is obtained are exceptional. The number of iterations depends on the 3BP/4BP case considered, on the boundary values, on the time of flight, and, obviously, on the initial guess adopted.
	
\subsection{Transfers in the CR3BP}
The CR3BP is represented by neglecting the gravitational influence of the Sun, i.e., assuming $\mu_s=0$ in Eq. \ref{eq:1}. This model is considered in this subsection. In the CR3BP, this system has three main parameters related to the trajectory of the transfer: the position to apply the first impulse in the initial orbit, described by the parameter $\alpha$, the position at which the spacecraft applies the second impulse near the Moon, $\beta$, and finally, the transfer time $T$. In order to investigate the relations of all of these parameters with the best $\Delta V$, the following procedure is used. First, the parameter $T$ is fixed; thus, the dependency of $\alpha$ and $\beta$ on the minimum $\Delta V$ is to be found for a specific time of flight $T$. The minimum $\Delta V$ is numerically found by sweeping the parameters $\alpha$ and $\beta$ in their neighborhood one at a time and maintaining the lowest value among the ones numerically found. After that, the neighborhood in which each parameter is evaluated is decreased, and the process is repeated several times for all the parameters. The whole process is repeated until the change in the $\Delta V$ is less than $1 \times 10^{-3}$ m/s, and the change in the angles are less than $1 \times 10^{-3}$ rad. Although the position error of the solution obtained through TFC is always smaller than one meter (about five orders of magnitude smaller than one meter in most cases), the obtained solution is not necessarily the one with the lowest $\Delta V$. Thus, the main source of error in the procedure to find the parameters' relations is the convergence. We deal with a complex non-linear problem with multiple solutions, and the solution may converge to ones that are not of interest. Thus, although the process stops when the above changes in $\Delta V$ and angles are reached, it does not necessarily mean that this range is the error for the solution of interest, i.e., the minimum $\Delta V$ for that time of transfer $T$. On the other hand, the accuracy is good enough to investigate the parameters' variations as functions of the time of flight for the best $\Delta V$, as can be seen in the following results.
	
\begin{figure}[!ht]
	\centering
	\includegraphics[width=0.8\linewidth]{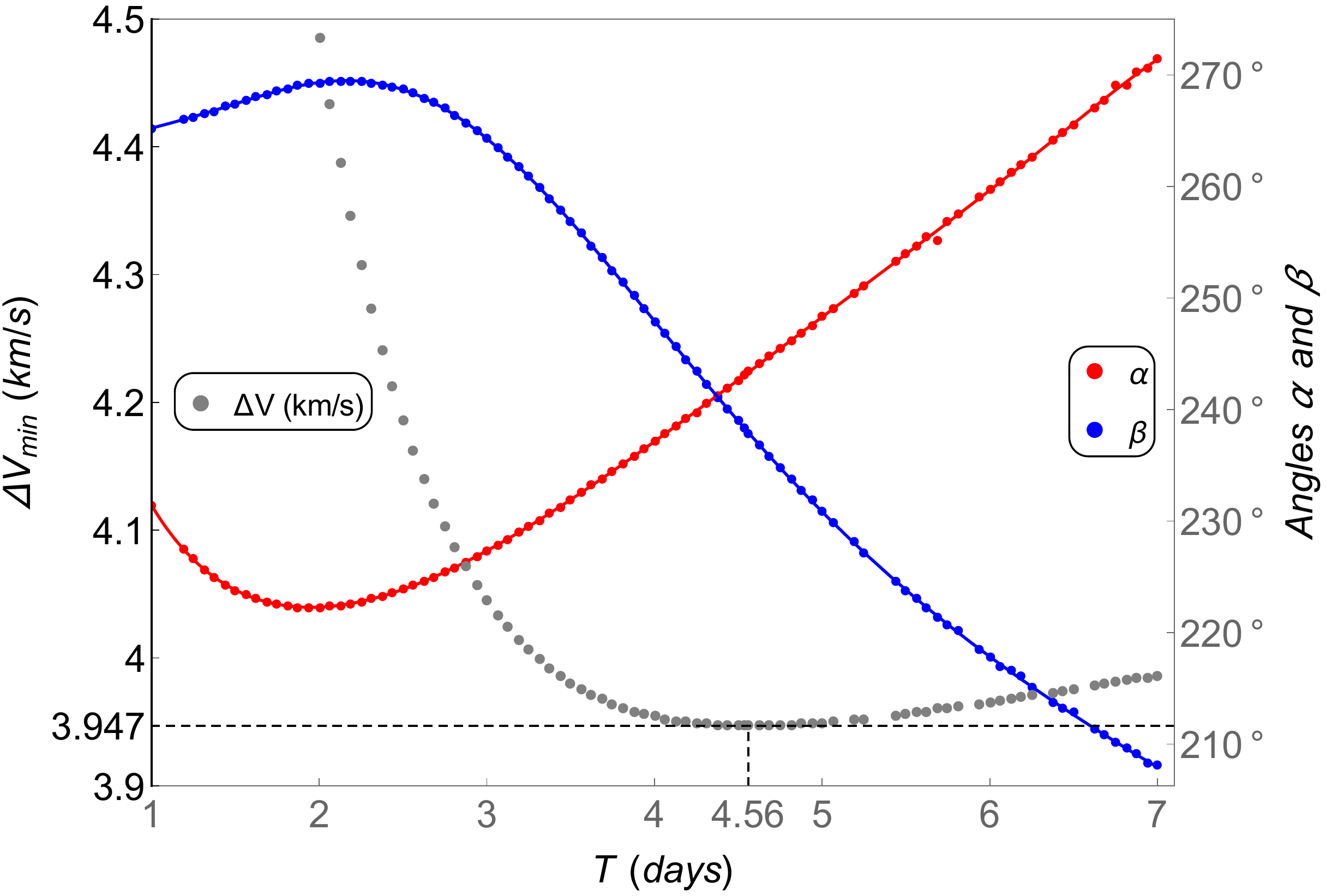}
	\includegraphics[width=0.8\linewidth]{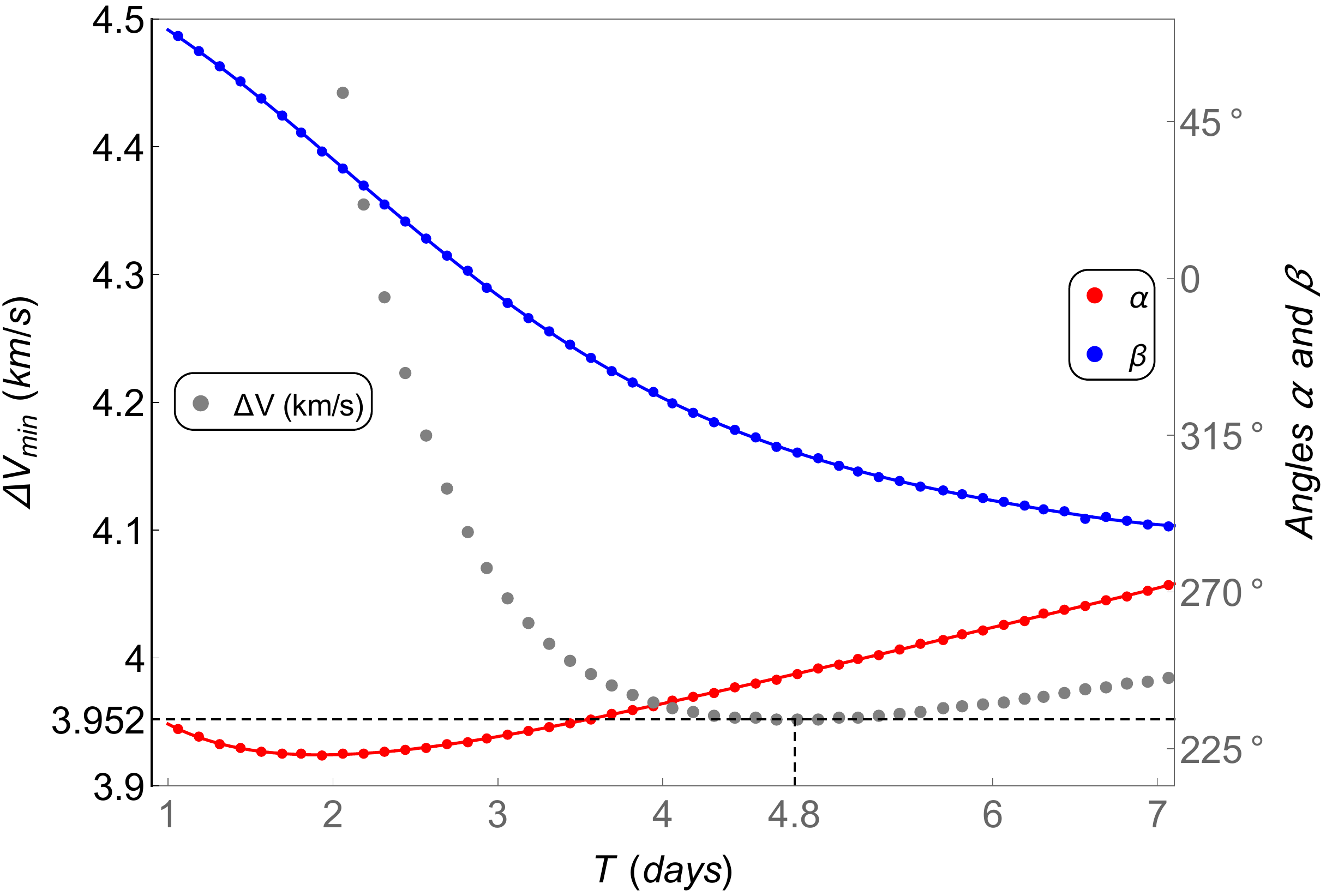}
	\caption{Relations between all of the parameters in CR3BP transfer: the angles $\alpha$ (in red) and $\beta$ (in blue) associated with the trajectory with the minimum $\Delta V$ found (in gray) as functions of the time of flight $T$. The dots are the values found in this research, while the curves are fits to these dots. The upper figure is the counter-clockwise case and lower one is the clockwise.}
	\label{fig:r1}
\end{figure}

\begin{figure}[!ht]
	\centering
	\includegraphics[width=0.47\linewidth]{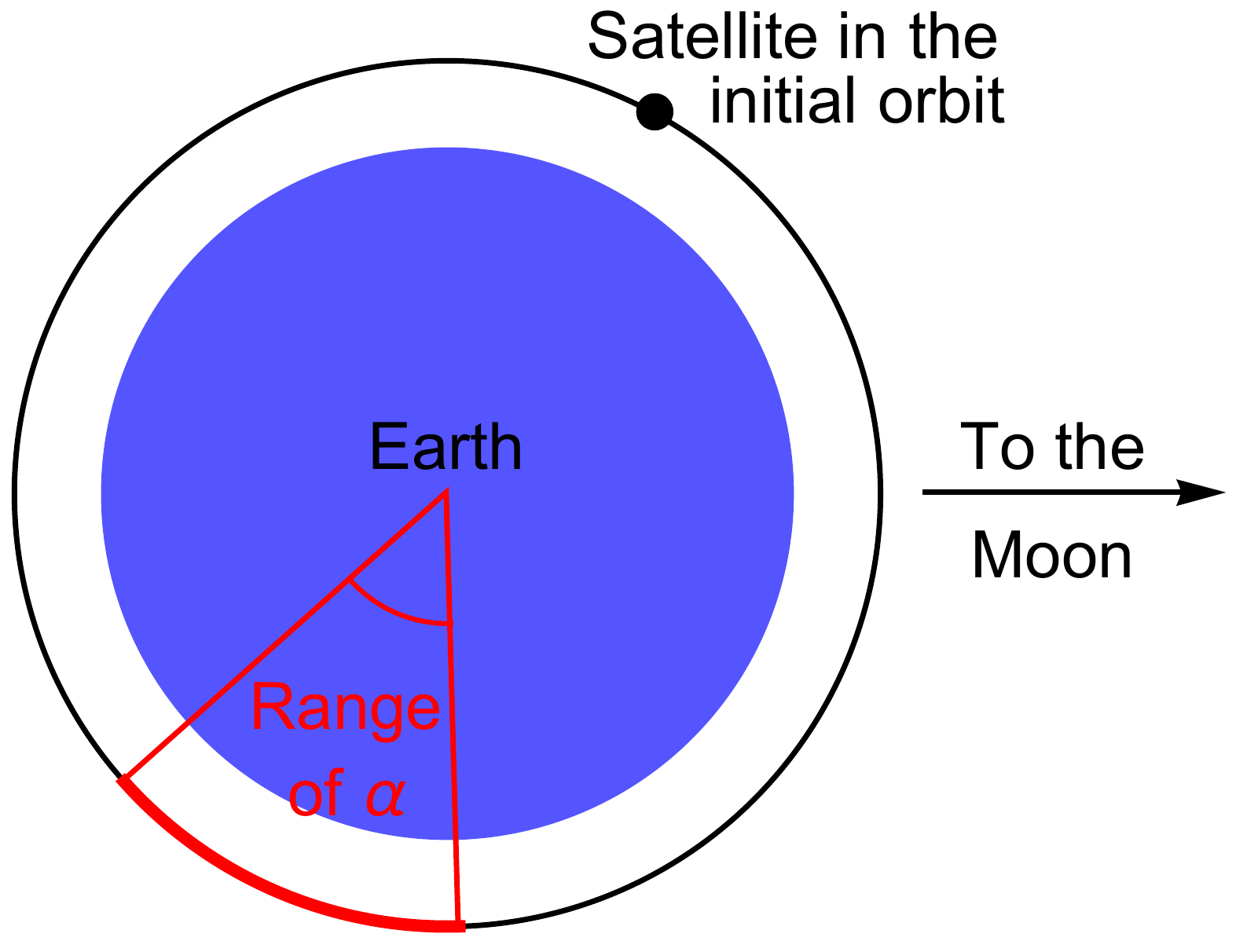}
	\includegraphics[width=0.47\linewidth]{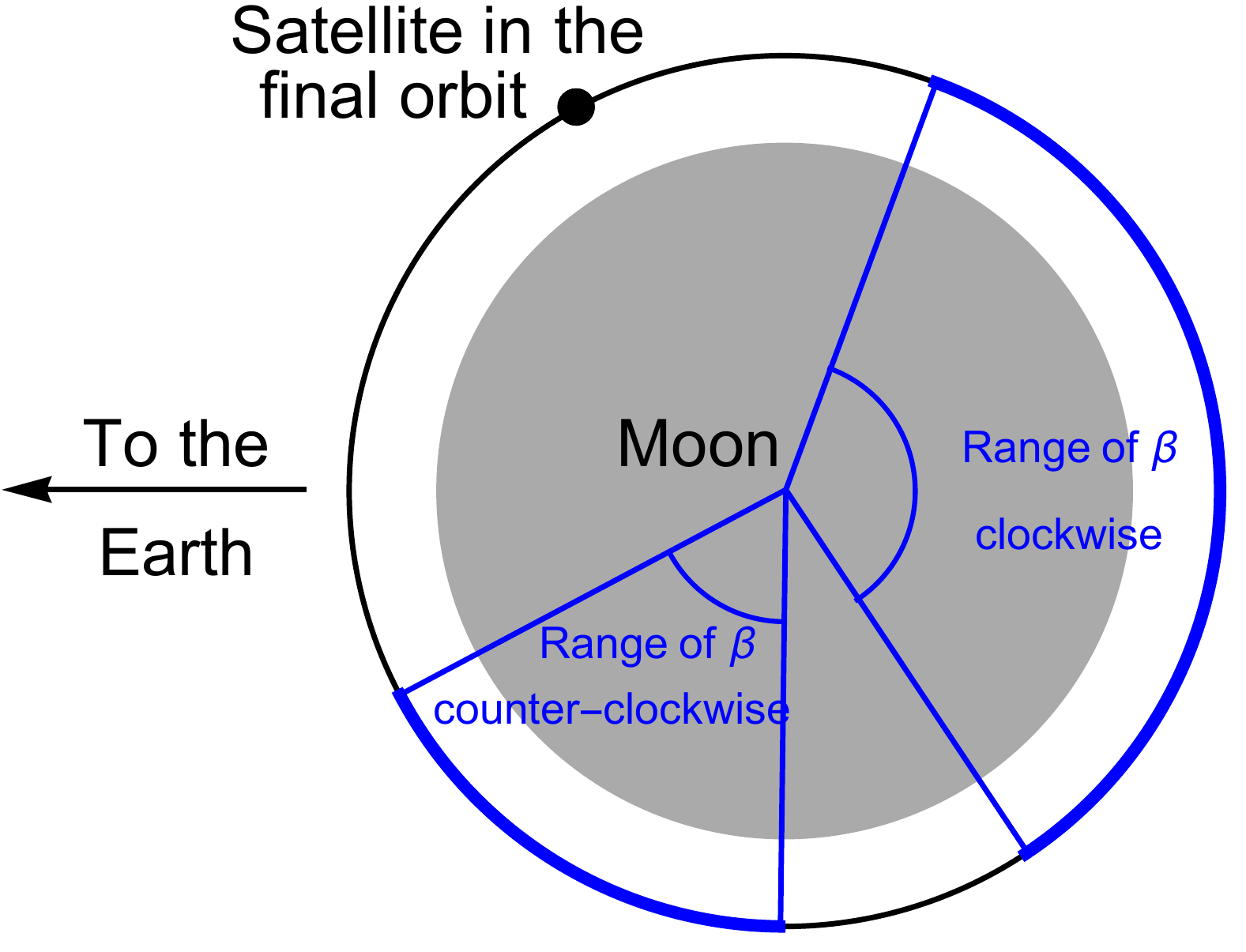}
	\caption{Range of the angles of departure $\alpha$ and arrival $\beta$ that minimizes the costs $\Delta V$ for the counter-clockwise ($\beta_{c-c}$) and clockwise ($\beta_{clockwise}$) cases for transfers from 1 to 7 days.}
	\label{fig:3bpbeta}
\end{figure}

Using the process described above, the trajectories with the lowest costs are found, and the parameters related to these trajectories are disclosed. The results can be seen in Fig. \ref{fig:r1}, where the dependencies of the parameters $\alpha$ (in red) and $\beta$ (in blue) on the time of flight $T$ are shown for a final counter-clockwise orbit (above) and a final clockwise orbit (below). The values of the transfer costs are shown in gray in km/s. The red and blue curves are curve fits for the dots using tenth order Hermite polynomials. The parameter $\alpha$ decreases as $T$ increases for short-time transfers up to two days and increases with $T$ after this value in a range from 222$^{\circ}$ to 272$^{\circ}$. This behavior for $\alpha$ is identical for both the counter-clockwise and the clockwise cases. In fact, there is no considerable difference in the values of this parameter for both cases. In the counter-clockwise case, the parameter $\beta$ increases as $T$ increases for $T$ up to almost three days and decreases with $T$ after this value in a range from 270$^{\circ}$ to 208$^{\circ}$. Considering the clockwise case, the parameter $\beta$ always decreases when $T$ increases in a different range from about 70$^{\circ}$ to -70$^{\circ}$ (or 290$^{\circ}$). The best ranges for $\alpha$ and $\beta$ can be seen in Fig. \ref{fig:3bpbeta}. Using the procedure described above, all parameters are varied to search for the minimum $\Delta V$, including the time of flight $T$. The values of all parameters associated with the minimum $\Delta V$ are shown in Table \ref{tab:t1} for the counter-clockwise and clockwise cases.
	
	
\begin{table*}[!ht]
	\centering
	\begin{tabular}{|l|r|r|}
		Variables associated&\multicolumn{1}{c|}{Counter-clockwise}&\multicolumn{1}{c|}{Clockwise}\\
		\hline
		$\Delta V~(\text{m}/\text{s})$&$3946.93$&$3952.01$\\
		$\|\delta \B{V}_A\|~(\text{m}/\text{s})$&$3134.60$&$3137.32$\\
		$\|\delta \B{V}_B\|~(\text{m}/\text{s})$&$812.33$&$814.693$\\
		$T~(\text{days})$&$4.55395$&$4.7997$\\
		$\alpha~(\text{rad})$&$4.24587$&$4.30199$\\
		$\alpha$&$243.3^{\circ}$&$246.5^{\circ}$\\
		$\beta~(\text{rad})$&$4.15460$&$5.41481$\\
		$\beta$&$238.0^{\circ}$&$310.2^{\circ}$\\
		$P_E~(\text{m})$&$4.6 \times 10^{-4}$&$1.4\times 10^{-6}$\\
		$V_E~(\text{m}/\text{s})$&$2.9 \times 10^{-7}$&$9.6\times 10^{-10}$\\
		$\dot{x}|_{t=t_0}~(\text{m}/\text{s})$&$9745.19$&$10007.6$\\
		$\dot{y}|_{t=t_0}~(\text{m}/\text{s})$&$-4907.6$&$-4354.4$\\
		\hline
	\end{tabular}
	\caption{Values associated to the minimum $\Delta V$ found for the transfer in the CR3BP.}
	\label{tab:t1}
\end{table*}
		
\subsection{Transfers in the bi-planar bi-circular 4BP}
	
\begin{figure}[!htbp]
	\centering
	\includegraphics[width=0.8\linewidth]{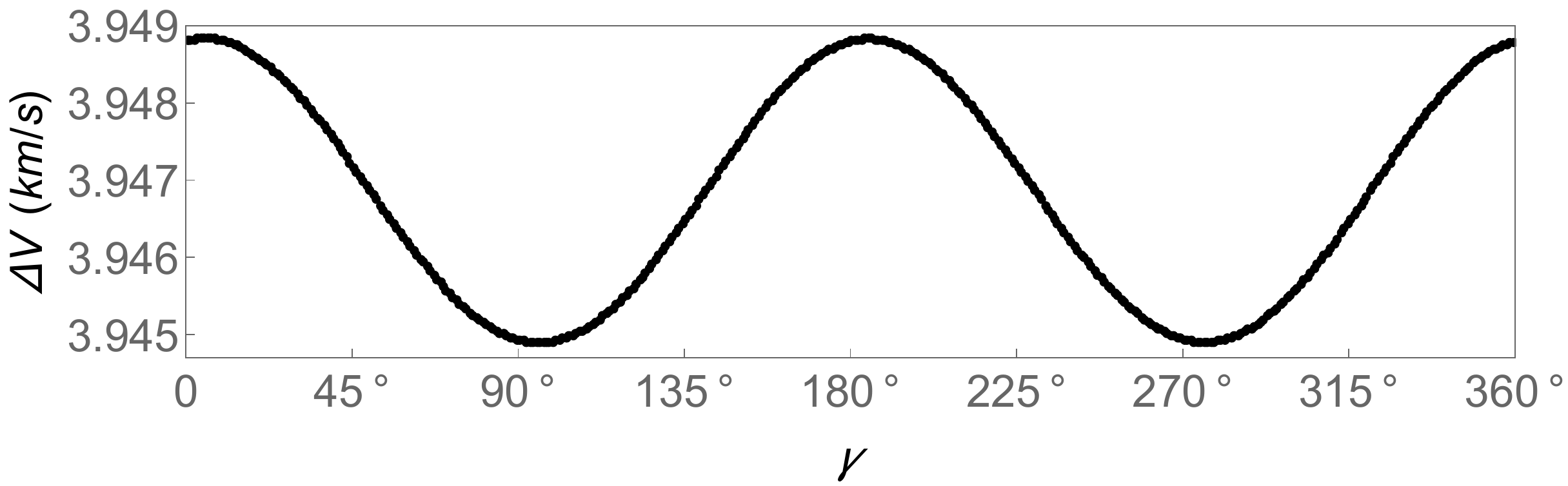}
	\caption{The $\Delta V$ cost as a function of the position of the Sun $\gamma$ for the parameters $\alpha$, $\beta$ and $T$ fixed in the values given in Table \ref{tab:t1}.}
	\label{fig:f2}
\end{figure}

\begin{figure}[!htbp]
	\centering
	\includegraphics[width=0.8\linewidth]{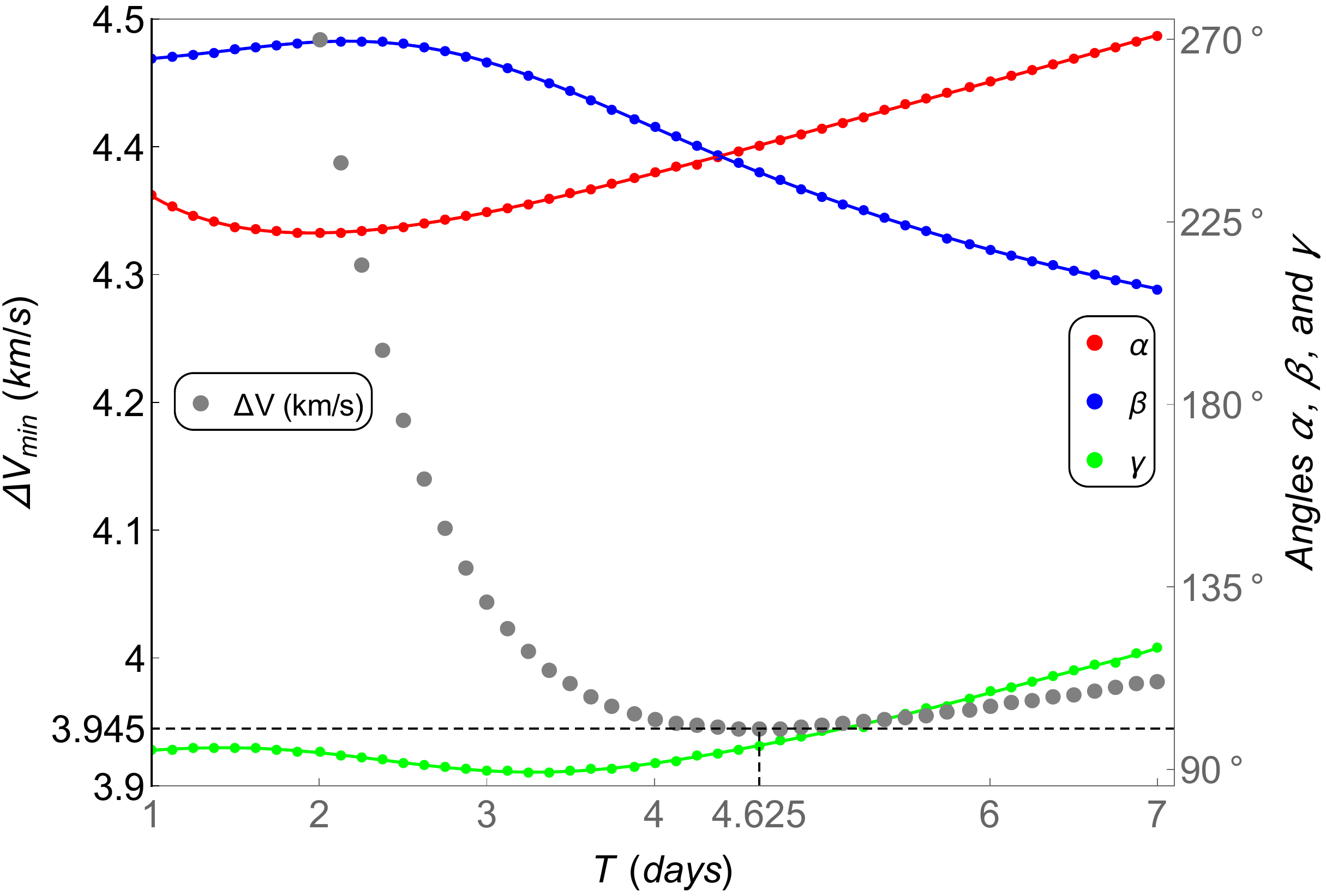}
	\includegraphics[width=0.8\linewidth]{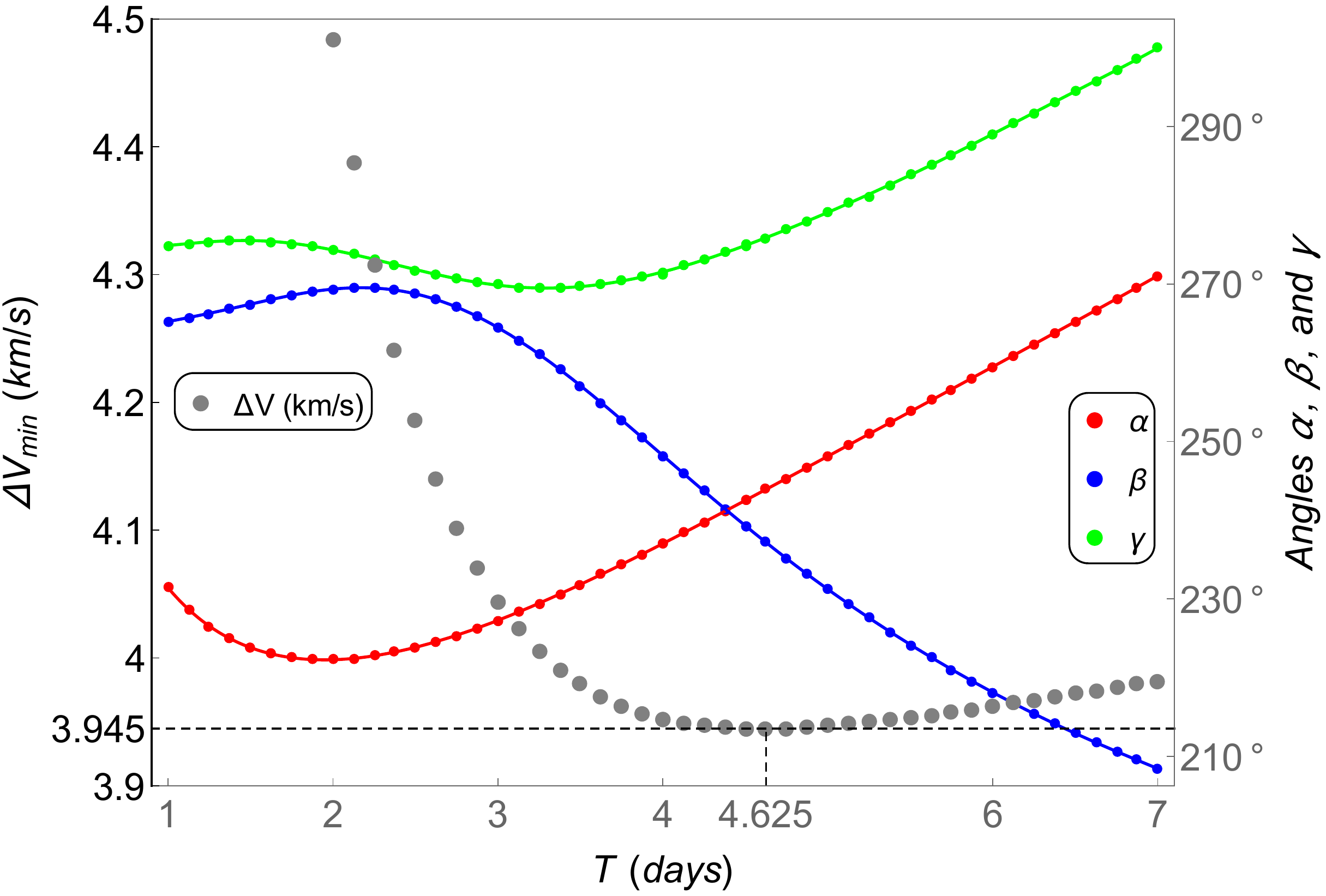}
	\caption{The counter-clockwise case: dependencies among all of the parameters in 4BP transfer: the angles $\alpha$ (in red), $\beta$ (in blue), and $\gamma$ (in green) associated with the trajectory with the minimum $\Delta V$ found (in gray) as functions of the time of flight $T$.}
	\label{fig:r3}
\end{figure}

\begin{figure}[!htbp]
	\centering
	\includegraphics[width=0.8\linewidth]{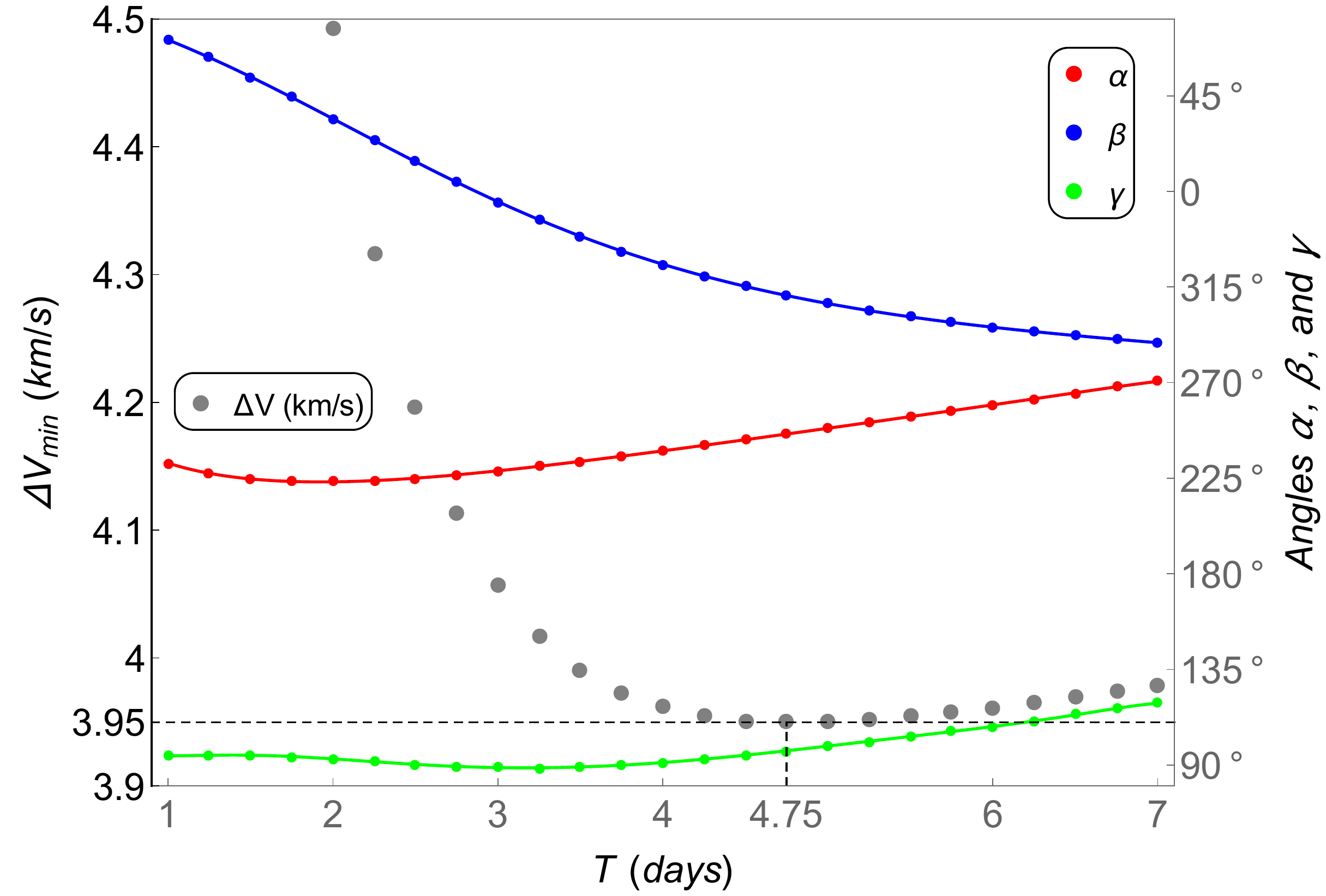}
	\includegraphics[width=0.8\linewidth]{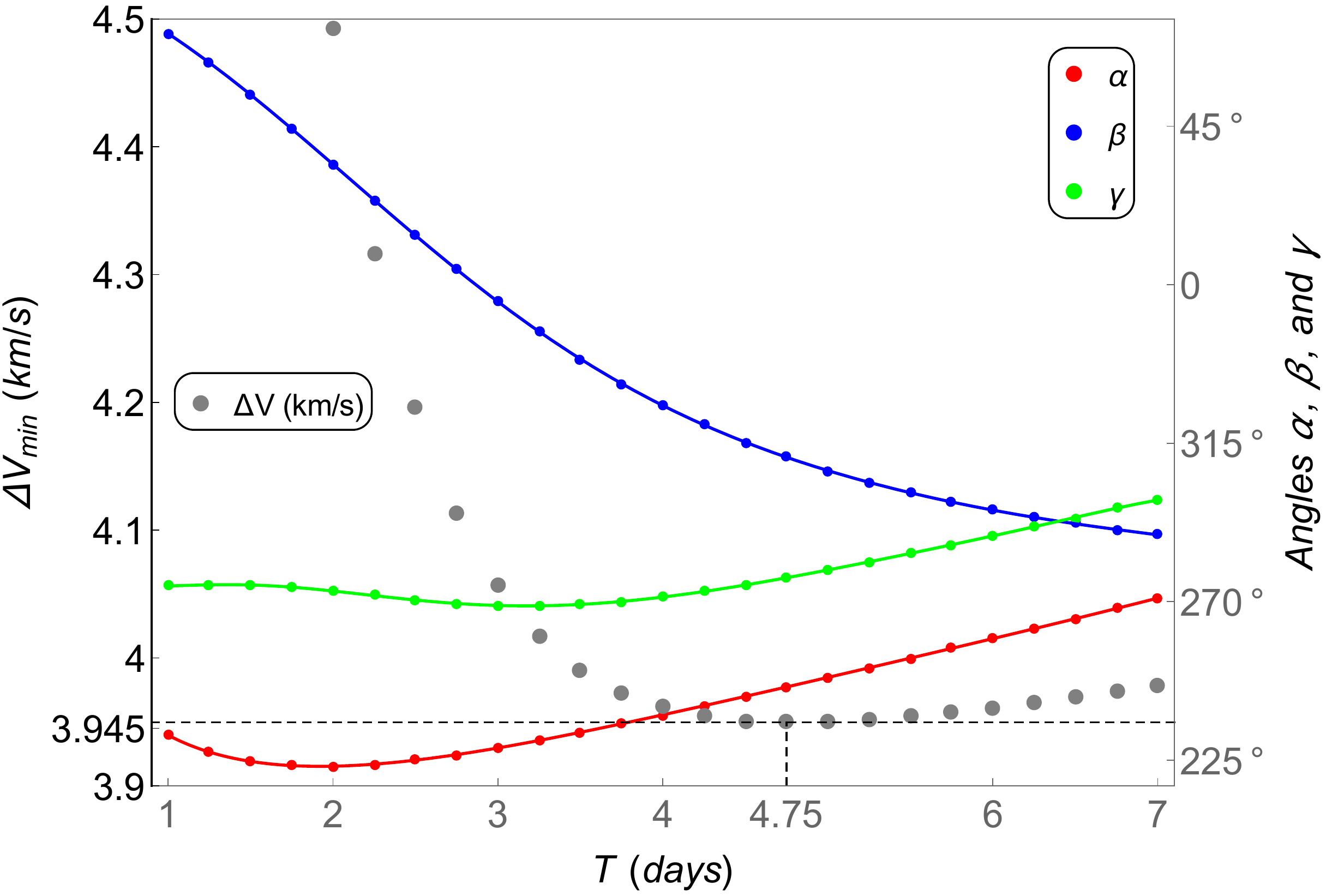}
	\caption{The clockwise case: dependencies among all of the parameters in 4BP transfer: the angles $\alpha$ (in red), $\beta$ (in blue), and $\gamma$ (in green) associated with the trajectory with the minimum $\Delta V$ found (in gray) as functions of the time of flight $T$.}
	\label{fig:r4}
\end{figure}

\begin{figure}[!ht]
	\centering
	\includegraphics[width=0.47\linewidth]{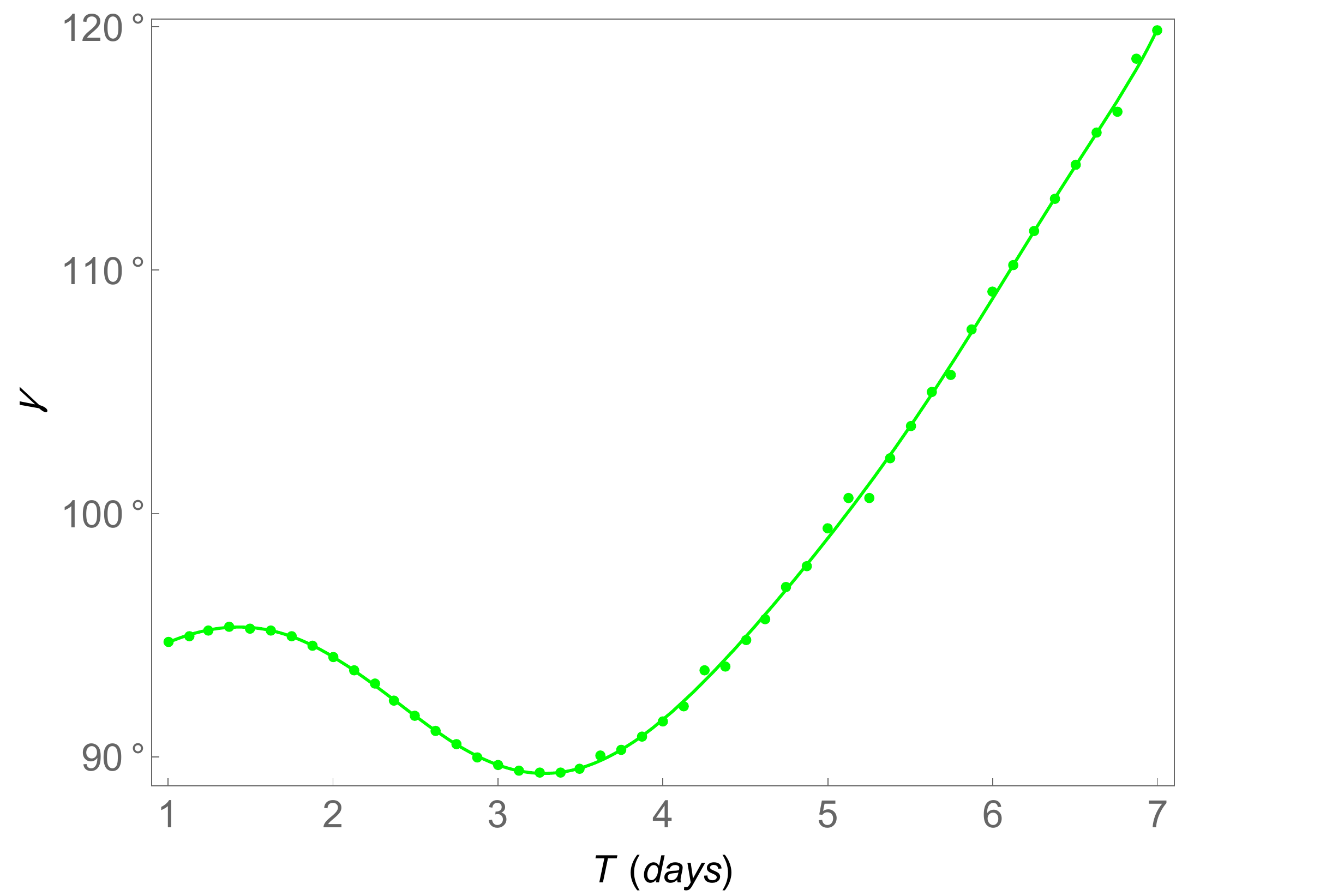}
	\caption{Values of $\gamma$ that minimizes the costs $\Delta V$ for the counter-clockwise transfer.}
	\label{fig:4bpgammaccl}
\end{figure}

\begin{figure}[!htbp]
	\centering
	\includegraphics[width=0.47\linewidth]{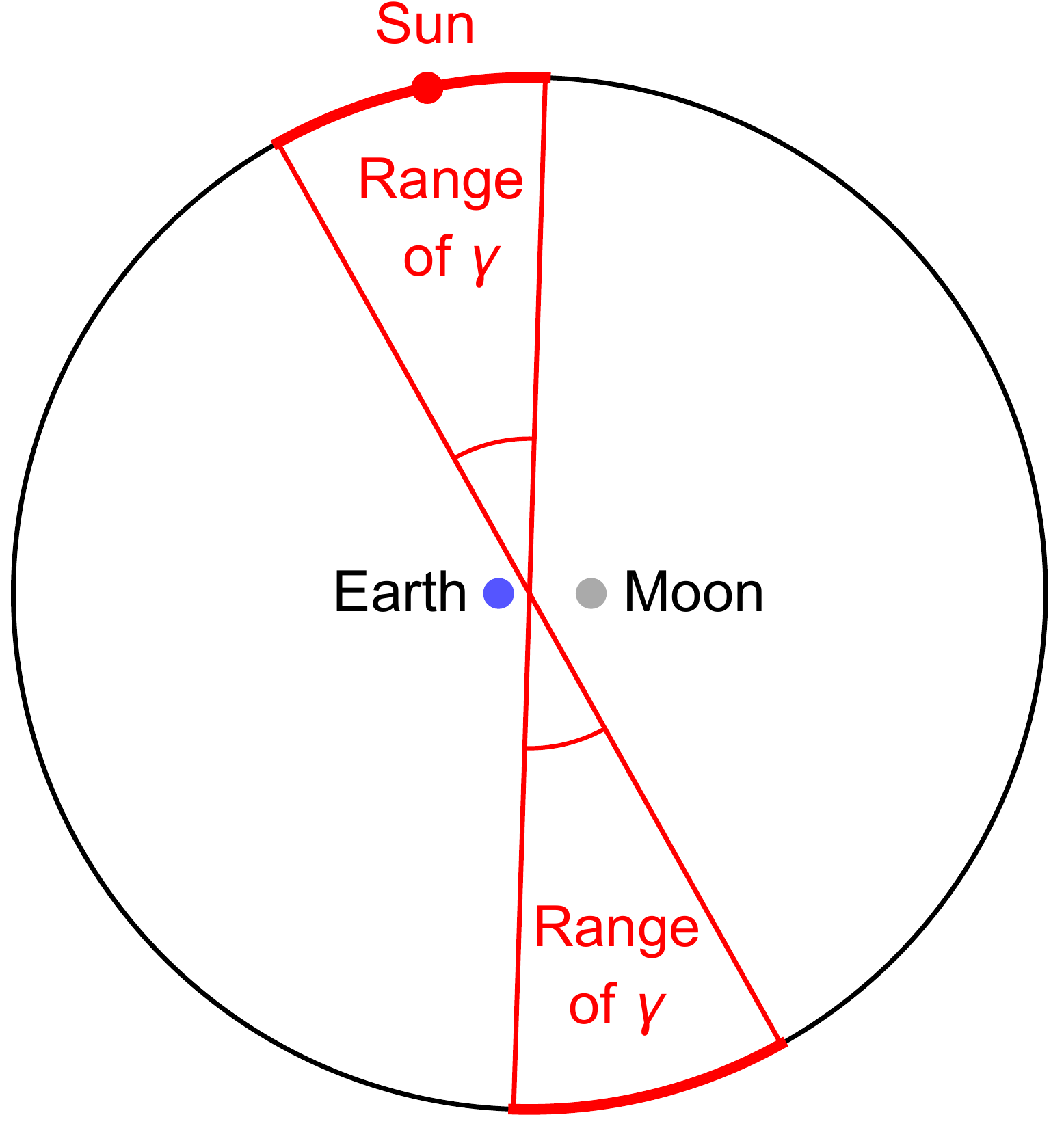}
	\caption{Range of the relative position of the Sun that minimizes the costs $\Delta V$ with respect to the Earth and Moon for transfers from 1 to 7 days.}
	\label{fig:4bpgamma}
\end{figure}

In order to analyze the influence of the Sun, the complete system given by Eq. \eqref{eq:1} is considered with the parameters given in Table \ref{tab:1}. To analyze the influence of the Sun over the trajectory related to the minimum $\Delta V$ for the previous CR3BP, the parameters $\alpha$, $\beta$, and $T$ shown in Table \ref{tab:t1} are fixed, and the angle related to the position of the Sun is varied. The cost $\Delta V$ as a function of $\gamma$ is shown in Fig. \ref{fig:f2}. Note that the $\Delta V$ changes by about 2 m/s due to the Sun's influence over the transfer compared with the CR3BP. There are two $\Delta V$ local minima for $\gamma$ around $95^{\circ}$ or $275^{\circ}$. The dependencies of all the parameters $\alpha$, $\beta$, and $\gamma$ on the time of flight $T$ are shown in Fig. \ref{fig:r3} for the local minimum of $\gamma$ around $95^{\circ}$ (upper) and for the local minimum of $\gamma$ around $275^{\circ}$ (lower): both for the counter-clockwise case. The clockwise case is shown in Fig. \ref{fig:r4} for the same two best positions of the Sun. The dots correspond to the found minima, and the curves represent fitting to the points using tenth order Hermite polynomials. The values related to the lowest $\Delta V$ found in this research for the bi-planar, bi-circular 4BP are shown in Table \ref{tab:t2}. The values of $\alpha$ and $\beta$ that produce the lowest $\Delta V$ differ by up to 0.7$^{\circ}$ from the CR3BP: this can be seen by comparing the values reported in Tables \ref{tab:t1} and \ref{tab:t2}.  The Sun's influence changes the cost of the transfers by about 2 m/s, depending on the Sun's position. There are two minima and two maxima separated by a 180$^{\circ}$ phase with respect to the position of the Sun, $\gamma$: the position of the Sun for the lowest $\Delta V$ is about 96$^{\circ}$ or 276$^{\circ}$ (for $T\approx4.6$ days), but the best position of the Sun can vary by up to 120$^{\circ}$ or 300$^{\circ}$ for longer transfers times of about seven days. The variation of $\gamma$ for the counter-clockwise case shown in Fig. \ref{fig:r3} is also shown in an isolated form in Fig. \ref{fig:4bpgammaccl} for a better visualization of the range of $\gamma$. The range of the values of $\gamma$ that minimizes the fuel cost is shown in Fig. \ref{fig:4bpgamma}.
	
\begin{figure}[!htbp]
	\centering
	\includegraphics[width=0.7\linewidth]{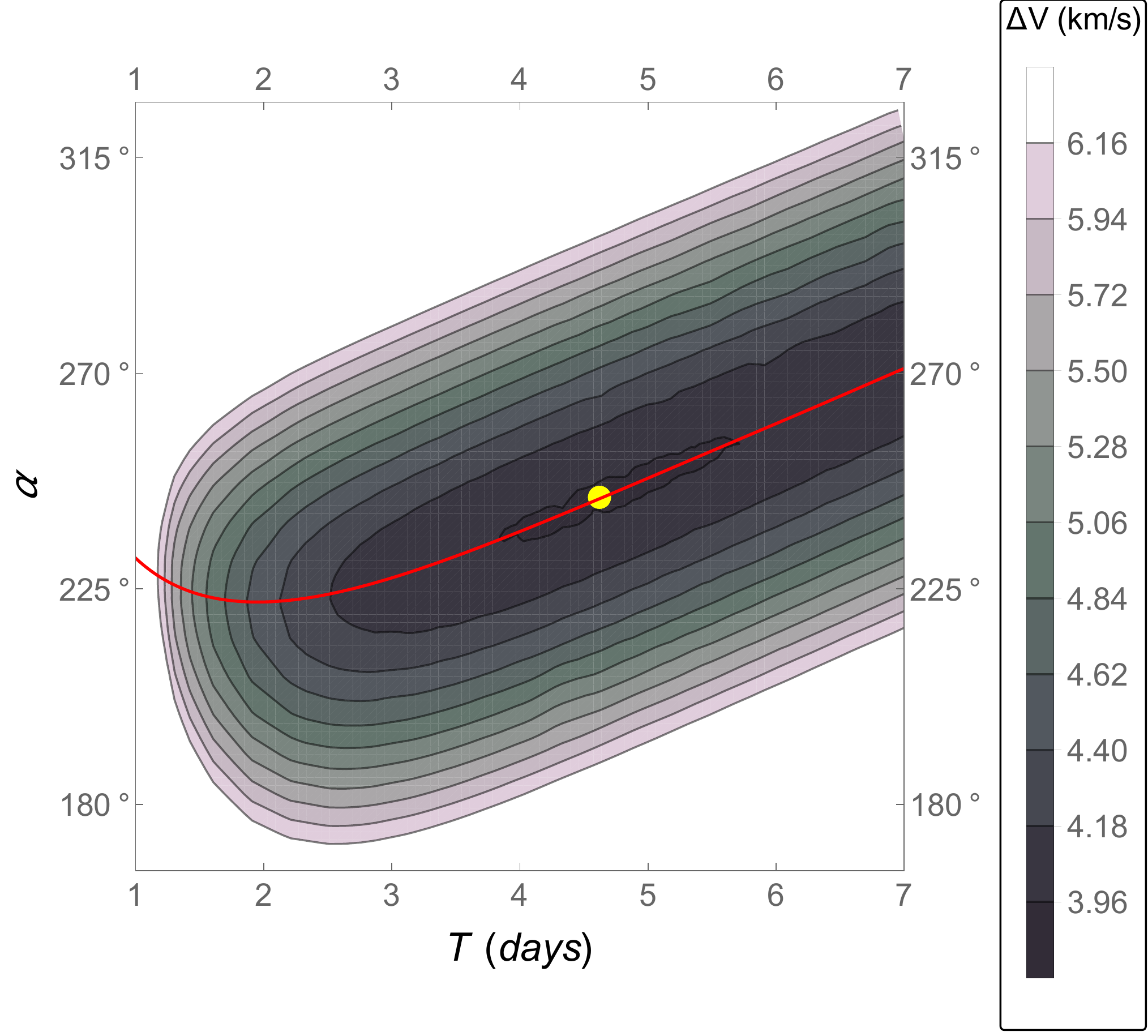}
	\caption{The cost $\Delta V$ as function of $\alpha$ and the total time of flight $T$. The yellow dot represents the pair $\alpha$ and $T$ for the minimum $\Delta V$.}
	\label{fig:dvat}
\end{figure}

\begin{figure}[!htbp]
	\centering
	\includegraphics[width=0.7\linewidth]{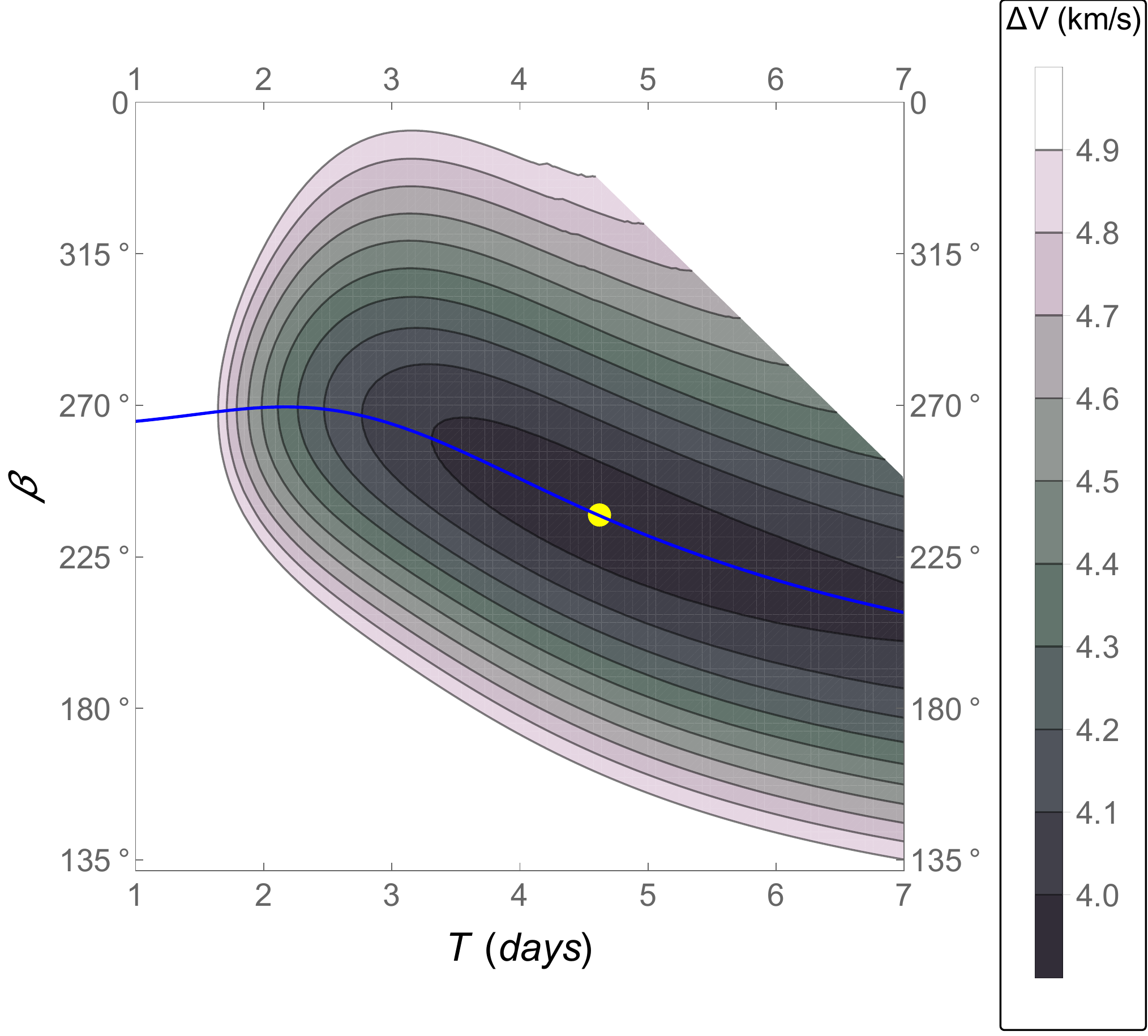}
	\caption{The counter-clockwise case: the cost $\Delta V$ as function of $\beta$ and the total time of flight $T$. The yellow dot represents the pair $\alpha$ and $T$ for the minimum $\Delta V$.}
	\label{fig:dvbt}
\end{figure}

\begin{figure}[!htbp]
	\centering
	\includegraphics[width=0.7\linewidth]{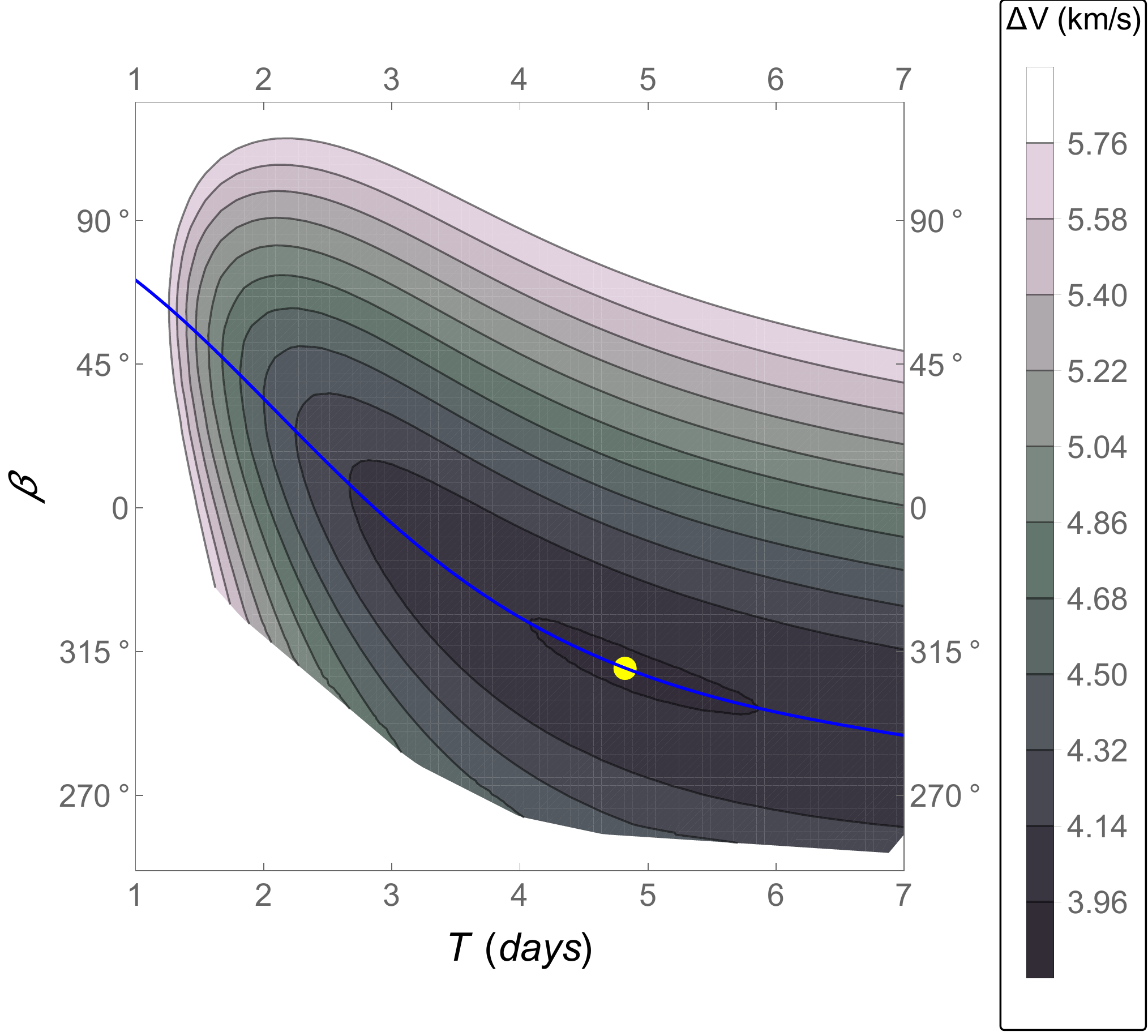}
	\caption{The clockwise case: the cost $\Delta V$ as function of $\beta$ and the total time of flight $T$. The yellow dot represents the pair $\alpha$ and $T$ for the minimum $\Delta V$.}
	\label{fig:dvbtclock}
\end{figure}

Since the values of $\alpha$, $\beta$, $\gamma$, and $T$ are disclosed in Figs. \ref{fig:r3} and \ref{fig:r4} for the lowest $\Delta V$, the effects of the variations of pairs of these parameters can now be investigated. The first analysis is on the pair $\alpha$ and $T$. In this case, the values of $\beta$ and $\gamma$ are constrained using the curve given by the Hermite polynomials fits of the dots in Fig. \ref{fig:r3} (upper). The cost $\Delta V$ as a function of $\alpha$ and $T$ is shown in Fig. \ref{fig:dvat}. The red curve in this figure is the same red curve of Fig. \ref{fig:r3} (upper), and it shows the values of $\alpha$ as a function of $T$ for the minimum $\Delta V$.
Note that the data shown in Fig. \ref{fig:dvat} is obtained for the counter-clockwise case, but the pattern of $\alpha$ is not changed for the clockwise case.
The second analysis is on the pair $\beta$ and $T$. In this case, the values of $\alpha$ and $\gamma$ as functions of $T$ are given by the Hermite polynomial fits shown in Fig. \ref{fig:r3} (upper). In this second case, the cost $\Delta V$ as a function of the pair $\beta$ and $T$ is shown in Fig. \ref{fig:dvbt}. The blue curve represents the values of $\beta$ for the minimum $\Delta V$, and it is the same curve shown in Fig. \ref{fig:r3} (upper). Note that this analysis is also done for the counter-clockwise case.
The third analysis shows the cost $\Delta V$ as a function of the pair $\beta$ and $T$, but this time for the clockwise case. In this case, the parameters $\alpha$ and $\gamma$ are given by the curve fits of Fig. \ref{fig:r4} (upper). The result is shown in Fig. \ref{fig:dvbtclock}. The blue curve represents the same blue curve of Fig. \ref{fig:r4} (upper), and it represents the values of $\beta$ that should be used to minimize the cost $\Delta V$. Note in Figs. \ref{fig:dvat} - \ref{fig:dvbtclock} that a small variation in the parameters $\alpha$ and $\beta$ leads to a large variation in the cost $\Delta V$.
	
	
\begin{table*}[!htbp]
	\centering
	\begin{tabular}{|l|r|r|}
		Variables associated&\multicolumn{1}{c|}{Counter-clockwise}&\multicolumn{1}{c|}{Clockwise}\\
		\hline
		$\Delta V~(\text{m}/\text{s})$&$3944.83$&$3949.73$\\
		$\|\delta \B{V}_A\|~(\text{m}/\text{s})$&$3134.41$&$3137.12$\\
		$\|\delta \B{V}_B\|~(\text{m}/\text{s})$&$810.421$&$812.61$\\
		$T~(\text{days})$&$4.625$&$4.81961$\\
		$\alpha~(\text{rad})$&$4.25717$&$4.30321$\\
		$\alpha$&$243.9^{\circ}$&$246.6^{\circ}$\\
		$\beta~(\text{rad})$&$4.13962$&$5.4084$\\
		$\beta$&$237.2^{\circ}$&$309.9^{\circ}$\\
		$\gamma~(\text{rad})$&$1.66965$&$1.69787$\\
		$\gamma$&$95.7^{\circ}$&$97.3^{\circ}$\\
		$P_E~(\text{m})$&$3.2 \times 10^{-4}$&$9.6\times 10^{-6}$\\
		$V_E~(\text{m}/\text{s})$&$2.1 \times 10^{-7}$&$6.6\times 10^{-9}$\\
		$\dot{x}|_{t=t_0}~(\text{m}/\text{s})$&$9799.8$&$10012.3$\\
		$\dot{y}|_{t=t_0}~(\text{m}/\text{s})$&$-4797.2$&$-4343.03$\\
		\hline
	\end{tabular}
	\caption{Values associated to the minimum $\Delta V$ found for the transfer using the bi-circular bi-planar R4BP.}
	\label{tab:t2}
\end{table*}
	
\subsection{Comparisons with the literature}
	
The minimum costs can be identified in the plane $T$-$\Delta V$ available in the literature. The data for the best $\Delta V$ and their respective time of flight $T$ available in the literature for a time of flight up to seven days are shown in Table \ref{tab:8}.
\begin{table*}[!ht]
	\centering
	\begin{tabular}{|l|r|l|l|}
		\multicolumn{1}{c|}{Model}&\multicolumn{1}{c|}{Literature reference}&\multicolumn{1}{c|}{$T$ (days)}&\multicolumn{1}{c|}{$\Delta V$ (m/s)}\\
		\hline
		PCR3BP&\cite{doi:10.2514/1.7702}&$3.1$&$4042$\\
		PCR3BP&\cite{doi:10.2514/1.7702}&$3.2$&$4034$\\
		PCR3BP&\cite{doi:10.2514/1.7702}&$3.3$&$4022$\\
		PCR3BP&\cite{doi:10.2514/1.7702}&$3.4$&$4007$\\
		PCR3BP&\cite{YAGASAKI2004313}&$3.88662$&$3962.84$\\
		Hohmann-like&\cite{topputo2013optimal}&$5$&$3954$\\
		PCR3BP&\cite{YAGASAKI2004313}&$4.56841$&$3951.57$\\
		PCR3BP&TFC counter-clockwise&$4.5625$&$3946.93$\\
		bi-planar/circular R4BP&\cite{Yagasaki2004}&$4.59446$&$3949.53$\\
		bi-planar/circular R4BP&\cite{topputo2013optimal}&$4.6$&$3944.8$\\
		bi-planar/circular R4BP&TFC counter-clockwise&$4.625$&$3944.83$\\
		\hline
	\end{tabular}
	\caption{The best $\Delta V$ found in the literature for a short transfer time (up to seven days) from the Earth to the Moon compared with the results for the best counter-clockwise and clockwise cases found in this paper. The altitudes of the orbits around the Earth and the Moon are $167$ km and $100$ km, respectively.}
	\label{tab:8}
\end{table*}
In the Hohmann-like transfer, the spacecraft is under the influence of only the Earth or only the Moon gravitational attraction in the trajectory's initial and final parts, respectively. This type of transfer is largely used to compare fuel cost efficiency $\Delta V$ and time of flight \cite{pernicka95,doi:10.2514/3.21079,circi01,perozzi08,topputo2013optimal,oshimatop19}. The PCR3BP is used with hybrid genetic algorithms and deterministic simplex methods to solve the boundary-value problem and obtain the data in \cite{doi:10.2514/1.7702}. The PCR3BP is also used with a computer software called \textit{AUTO} to obtain the data in \cite{YAGASAKI2004313}. The same software is used to obtain optimal values for the same transfer based on the co-circular, co-planar restricted four-body problem in \cite{Yagasaki2004}. Topputo`s data are obtained from Pareto optimal solutions using the direct transcription and shooting strategy through the bi-circular restricted four-body problem \cite{topputo2013optimal}.	The gray dots shown in Figs. \ref{fig:r3} and \ref{fig:r4} form the same Pareto optimal solutions given near the point (i) in \cite{topputo2013optimal}. 
	
Note from Table \ref{tab:8} that the TFC counter-clockwise for the PCR3BP is $4.64$ m/s better than the best case available in the literature. In this case, the velocity of the spacecraft at the point B before the impulse is $\{2068.97, -1290.77\}\T$ m/s while the velocity after the impulse - according to section \ref{sec:transfer} - is $\B{V}_{Bf}=\{1379.77, -860.796\}\T$ m/s. The angle between these two velocities vectors is less than $10^{-4}$ rad. Note from section \ref{sec:transfer} that the velocity $\B{V}_{Bf}$ is tangential to the orbit, because the final orbit is circular in the frame of reference fixed in the Moon. Hence, the velocity at point B given by TFC (before the impulse) is also tangential to the orbit. In fact, the obtained solutions for the minimum $\Delta V$ tend to converge to initial and final tangential orbits, and this result agrees with \cite{doi:10.2514/2.5062}. Note that the restriction that the initial and final velocities must be tangential to the initial and final orbits, respectively, is present in the methods to evaluate the $\Delta V$ in all the literature references shown in Table \ref{tab:8}. In contrast, the TFC method does not require this assumption.
	
\section{Conclusions}
	
This study takes advantage of TFC's ability to embed constraints into functionals and uses them to solve Two-Point Boundary Value Problems (TPBVP) appearing in perturbed orbit transfer problems. The TPBVP example selected is the Earth-to-Moon, two-impulse orbit transfer, where the third body influence moves from being a small perturbation to the dominant force. The TFC functionals analytically satisfy the boundary conditions and turn the problem into an unconstrained one. This allows us to solve the problem by nonlinear least-squares with automatic differentiation and a just-in-time (JIT) compiler, rather than shooting methods adopted by competing approaches. The initial guess is a spiral trajectory, and it allows us to obtain fast convergence with residual error on the order of $10^{-15}$. This fast procedure can be used to quickly build non-Keplerian pork-chop plots that can be used for analysis and/or preliminary mission design.
	
The dynamics of this TPBVP  has, in general, multiple solutions that are associated with local minima. This means that, for assigned boundary conditions, there might be several different solutions satisfying the equations of motion. Therefore, the convergence to the desired minimum $\Delta V_{\rm tot}$ solution depends on the initial guess. On the other hand, since the convergence is almost always obtained, the method can be used to find all sub-optimal solutions. In general, no more than five distinct solutions for assigned boundary conditions are found. The longer the transfer time, the more solutions there are. However, when all the solutions have been identified, the selection of the optimal is trivial.
	
This fast approach allowed us to analyze the behavior of the $\Delta V_{\rm tot}$ as a function of all four main parameters: the time of flight, the relative Sun direction, and the locations where the two impulses are applied. The analysis is performed to investigate the total costs around the optimal value of the parameters to quantify the effective contribution. This way, by discretization, pork-chops plots can be built for any parameter pair. Summarizing, the TFC method to solve this  astrodynamics TPBVP allowed us:
\begin{itemize}
\item to quickly find all distinct solutions;

\item to build perturbed trajectory pork-chops to understand the cost variations as a function of the parameters' variations.
\end{itemize}
	
Specifically, two Earth-to-Moon trajectories, with transfer times up to seven days were analyzed for two target orbits around the Moon, one clockwise and one counter clockwise, respectively. The transfer minimum $\Delta V_{\rm tot}$ cost obtained in the bi-planar, bi-circular 4BP case coincides with the best values found in literature. In the PCR3BP case, the proposed TFC-based approach provides a slightly better solution than those found in literature.
	
\begin{acknowledgements}
	This work was supported by FAPESP - S\~ao Paulo Research Foundation through grants 2019/18480-5, 2018/07377-6 and 2016/24561-0, and by the NASA Space Technology Research Fellowship, Leake [NSTRF 2019] Grant \#: 80NSSC19K1152 and Johnston [NSTRF 2019] Grant \#: 80NSSC19K1149.
\end{acknowledgements}

\bibliographystyle{plain}       
\bibliography{ref1}

\end{document}